\documentclass[submit]{smj}

\usepackage[utf8]{inputenc}

\Author{M. Carmen Aguilera-Morillo\Affil{1} and Ana M. Aguilera\Affil{2}}
\AuthorRunning{M. Carmen Aguilera-Morillo and Ana M. Aguilera}

\Affiliations{

\item Department of Statistics,
      Escuela Polit\'ecnica Superior and UC3M-BS Santander Big Data Institute,
      Universidad Carlos III de Madrid,
      Madrid, Spain

\item Department of Statistics and O. R. and IEMath-GR,
      Facultad de Ciencias,
      Universidad de Granada,
      Granada, Spain

}   

\CorrAddress{M. Carmen Aguilera-Morillo,
             Department of Statistics,
             Escuela Polit\'ecnica Superior,
             Universidad Carlos III de Madrid,
             Avda. de la Universidad 30,
             28911, Madrid,
             Spain}
\CorrEmail{mariacarmen.aguilera@uc3m.es}
\CorrPhone{(+0034)\;916\;249\;179}
\CorrFax{(+0034)\;916\;249\;177}

\Title{Multi-class classification of biomechanical data: A functional LDA approach based on multi-class penalized functional PLS}
\TitleRunning{Multi-class classification of biomechanical data}

\Abstract{
A functional linear discriminant analysis (LDA) approach to classify a set of kinematic data (human movement curves of individuals performing different physical activities) is performed. Kinematic data, usually  collected in linear acceleration or angular rotation format, can be identified with functions in a continuous domain (time, percentage of gait cycle, etc). Since kinematic curves are measured in the same sample of individuals performing different activities, they are a clear example of functional data with repeated measures. On the other hand, the sample curves are observed with noise. Then, a roughness penalty might be necessary in order to provide a smooth estimation of the discriminant functions, which would make them more interpretable. Moreover, because of the infinite dimension of  functional data, a reduction dimension technique  should be considered. To solve these problems, we propose a multi-class approach for penalized functional partial least squares regression. Then LDA will be performed on the estimated functional PLS components.
This methodology is motivated by two case studies. The first study considers the linear acceleration recorded every two seconds in 30 subjects, related to three different activities  (walking, climbing stairs and down stairs). The second study works with the triaxial angular rotation, for each joint,  in 51 children when they completed a cycle walking under three conditions (walking, carrying a backpack  and pulling a trolley). A simulation study is also developed for comparing  the performance of the proposed functional LDA with respect  to the corresponding multivariate and non-penalized approaches.
}


\begin{document}

\maketitle

\section{Introduction}

Biomechanical gait data are commonly used to  differentiate between several gait pathologies or different physical activities. Different statistical methodologies for classification of gait data have been developed in the literature with this aim. Gait data are usually collected at different points in a continuous scale, such as time, so that they are generated from an underlying time-varying function.  In the applications that motivate this research, acceleration curves are measured at real time every two seconds, and rotation curves are measured in terms of the percentage of gait cycle. In spite of the continuous nature of kinematics curves, in many of the biomechanical studies the statistical analysis is made from a sample of summary measures for each curve such as minima, maxima, area under the curve, angles at heel strike, range of motion, or the timing of specific events \citep{Sadeghi1997,Sadeghi2003,Schmidt2010,Orantes2015}.

Currently, data science tools are being developed to build accurate classification and prediction models for gait research by using all the available information. One of them is Functional Data Analysis (FDA), that begins by reconstructing the true functional form of each sample curve from discrete sampling points that could  be different for different subjects. This way all the available information  about the sample curves is exploited to explain the movement in the statistical analysis. A detailed description of the most common FDA methodologies,  and interesting FDA applications with R and Matlab, are described in the pioneer books by \citet{RamsayI1997}, \citet{RamsayII} and \citet{RamsayIII}. A complete study of nonparametric FDA techniques can be seen in \citet{Ferraty2006}. Statistical inference for functional data has been studied in \citet{Horvath2012}. A more recent book is \cite{Kokoszka2017}, where new methods for FDA are collected (functional time series, spatial functional data, between others).

Biomechanical data are usually obtained over a number of discrete time points (snapshots) and assumed to be generated
by some underlying smooth function. Because of this, the steps typically used when FDA is
applied to the analysis of biomechanical data are: basis expansion representation, curve registration (time normalization or
landmark registration) and functional principal component analysis (FPCA)  (dimension reduction and variability explanation in terms of uncorrelated scalar variables).
The first step consists of obtaining smooth representations of the sample curves  in terms of a basis of functions (Fourier, B-splines, wavelets, etc). In the second step
registration is used to reduced phase variability between curves while preserving the individual curve's shape and amplitude. The effect of different registration methods on cyclical kinematic data is analyzed in \citet{Crane2010}. In the third step,
FPCA is successfully applied to explain the ways of variation in biomechanics of sports injuries and coordination in race-walking, jumping and running
\citep{Ryan2006,Harrison2007,Donoghue2008,Dona2009}. On the other hand, multivariate PCA on the raw discrete-time observations of the biomechanics curves has also been applied to provide biomechanics interpretation of the principal component scores and to discriminate between healthy and sick subjects
\citep{Deluzio1997,Daffertshofer2004,Milovanovic2012}.

This work is motivated by two different sets of biomechanical data: the human activity data and the gait data.
The first dataset is based on the human activity recognition collected by \citet{HumanActivity}. The original experiments were carried out with a group of 30 volunteers within an age bracket of 19-48 years. Each person performed three different activities wearing a smartphone on the waist (walking, walking upstairs and walking downstairs). Using its embedded accelerometer and gyroscope,
they captured 3-axial linear acceleration and 3-axial angular velocity at a constant rate of 50Hz every two seconds. The aim is to classify the curves of acceleration according to the categorical variable given by the three mentioned physical activities.
The gait dataset comes from a wide experimental study developed in the biomechanics
laboratories of the Sport and Health Institute of the University of
Granada (IMUDS).  In order to collect the data, twenty six reflective markers were placed on the skin of a total of 51 children  between
8 and 11 years old. The data were recorded by a 3D motion capture system meanwhile each subject completed a cycle walking over the platform
in three conditions (walking, carrying a backpack that weighs 20\% of the subject's weight and pulling a trolley that weighs 20\% of the subject's weight). For each subject, the 3-axial angular rotation were registered for
each join (ankle, foot progress, hip, knee, pelvis, thorax), and all conditions, in 101 equidistant points of the complete gait cycle. Now, the aim is to classify the rotation curves according to the three different load types.

Both case studies considered in this paper are clear examples of functional data  with repeated measures because the functional variables of interest (angular rotation and linear acceleration) are measured repeatedly on the same sample individuals under different experimental conditions. Therefore, the aim of this paper is to classify a set of functional data with repeated measures (kinematic curves) according to a categorical variable
with more than two categories (measurement conditions). In this paper, an extension of linear discriminant analysis to the case of functional data with repeated measurements will be considered. Alternatives classification approaches  based on logit regression were developed and successfully applied in different areas as environment, medicine, marketing, and so on \citep{Escabias2004,Escabias2005,Escabias2006,James2002,Delaigle2012a,AguileraPLS2013,Escabias2014}. An extension to functional
data of  nonparametric Bayes classifiers  based on
simple density ratios was proposed in \citep{Dai2017} for the case of binary classification.

Linear discriminant analysis (LDA) is a popular and consolidate methodology for classification and dimension reduction \citep{Fisher1936}.  So, LDA provides a sequence of linear combinations of the original  predictor variables (linear discriminants) that maximize the between-class variance  relative  to the within-class variance. An important problem is that LDA overfits the data when we have a large number of highly correlated predictor variables because the within-class covariance matrix is difficult to estimate. This is the case of functional data analysis that works with a large number of discrete-time observations for each sample curve. A way to solve this problem is to introduce some kind of regularization in the estimation of the  covariance matrix \citep{Friedman1989,Frank1989,Hastie1995}.  A general review of regularized techniques in discriminant analysis was developed in \citet{Mkhadri1997}. A Bayesian approach for Fisher's discriminant analysis of stochastic process was introduced in \citet{Shin2008}. An alternative solution,  for the case of irregularly sampled curves, is based on using as predictors a spline basis expansion  of the discrete-time observations  and  assuming Gaussian distribution  of the vector of basis coefficients with common covariance matrix for all classes, by analogy with LDA \citep{James2001}. This functional LDA approach was used to assessment of embryonic growth \citep{Bottomley2009}. Other way of solution consists of projecting the predictor vectors (multivariate analysis) or curves (functional data analysis)  onto a finite dimensional basis so that LDA, or any other discrimination procedures, can then be used taking the resulting basis coefficients as predictors.
A common approach is based on the representation of the original predictors by a small set of orthogonal scalar variables given by principal component analysis (PCA) or partial least squares (PLS) regression. Both approaches are commonly used in Chemometrics to discriminate between essential properties of substances from discrete measurements of the near-infrared (NIR) spectra. PCA was applied in \citet{Sato1994} to classify NIR spectral data of vegetable oils. PLS regression was performed in \citet{Li2000}  to discriminate edible fats and oils by Fourier transform NIR spectroscopy. Multivariate classification techniques such as LDA, quadratic
discriminant analysis (QDA) and logit regression  were applied in \citet{Indahl1999}. In the functional data context, reduction dimension techniques and regularization can be jointly used to get good classification rates and accurate interpretation of the results.
The functional LDA approach presented in this paper is inspired in this research line.

Functional PCA and functional PLS regression were introduced as natural extensions  of their multivariate counterparts to solve the problems of high dimension and multicollinearity associated with the scalar-on-function
linear model \citep{Deville1974,Dauxois1982,Ocana1999,Ocana2007,Preda2005,AguileraPLS2016}.
Both methodologies were compared on different simulated data sets concluding
that they have similar forecasting performance, but the estimated parameter function provided
by functional PLS regression is more accurate and needs fewer components
\citep{Reiss2007,AguileraPLS2010,Delaigle2012b,Delaigle2012c}. In addition, the performance of PCA for discrimination may not be optimal, because
PCA only identify gross variability, and then it is not capable to distinguish the between-groups and within-groups variability. In that sense, several authors proposed to use partial least squares (PLS) regression for dimension reduction, as for example \citet{Barker,Tenenhaus2003} in the multivariate context and \citet{Preda2007} in the functional context. 

From a  methodological point of view, the originality of our work lies, on the one hand, in proposing a penalized spline estimation of the PLS components to solve the problem of lack of smoothness, and on the other hand, in considering a multi-class approach for PLS regression based on the split-up variation to solve the problem of repeated measures. Then, classic LDA  is carried out on the estimated PLS components. To the best of our knowledge, it is the first approach in the literature for functional linear discriminant analysis with repeated measures based on functional PLS.
Then, in order to classify gait curves according to the kind of activity and take into account both, the lack of smoothness in the data and the correlation between repeated measurements, a new multi-class functional linear discriminant analysis  based on penalized functional PLS regression for repeated measures is proposed in Section \ref{multi-class}. The classification results of the considered multi-class functional LDA-PLS approach on simulated data and both biomechanical data sets are presented and discussed in Sections \ref{sim} and \ref{results}. Conclusions are finally included in Section \ref{conclus}.

\section{Multi-class classification of functional data}
\label{multi-class}
Let us consider a categorical response variable $Y=\{1,\ldots,K\},$ with $K$ $(K>2)$ categories and a functional predictor $X=\{X(t):\;
 t\in T\}$ (continuous and second order stochastic process whose sample paths belong to the Hilbert space of square-integrable functions $L_{2}(T)$).
 Then, the aim is to predict $Y$ from $X(t).$ That means to classify the sample curves according to the groups defined by the categories of $Y.$

The aim of functional linear discriminant analysis (FLDA) is to find linear combinations
\begin{equation}
\label{flda}
\Phi_{i}(X)=\alpha_{i} + \int_{T} X(t)\beta^{i}(t)dt, \quad \{i=1,\ldots,K-1\},
\end{equation}
so that the between class variance is maximized with respect to the total variance
$$
max_{\beta} \frac{\mathrm{V}\left(\mathrm{E}[\Phi(X) |Y] \right)}{\mathrm{V}\left(\Phi(X)\right)}.
$$

Due to the high dimension of $X(t)$ (the number of observed variables is usually larger than the number of sample observations), the estimation of the discriminant  functions is an ill-posed problem  because  the standard estimate for the within-class covariance matrix is singular so that it is impossible to apply the usual classification rule. This problem is solved in this paper by using a dimension reduction approach based on functional PLS regression \citet{Preda2007}. Then, the multi-class classification problem from repeated measures is solved by using the following two-step algorithm:
\begin{itemize}
 \item[Step 1.] Dimension reduction by penalized and non-penalized multi-class functional PLS regression (for repeated measures)
    of the random vector $\tilde{\pmb{Y}}=(\tilde{Y}_{1},\ldots,\tilde{Y}_{K-1})$ on the functional predictor $X(t)$, with $\{\tilde{Y}_{i} \in \{0,1\}: i=1,\ldots,K-1\}$ being the dummy variables associated with the categorical response Y.
    \item[Step 2.] Functional LDA of Y on a reduced set of functional PLS components estimated in the above step.
\end{itemize}

\subsection{Multi-class functional PLS regression}
\label{pls}
In order to reduce the dimension, partial least squares (PLS) regression of the vector $\tilde{\bf{Y}}=(\tilde{Y}_{1}, \dots ,\tilde{Y}_{K-1})$ on the functional predictor $X(t),$ with $\{\tilde{Y}_{i} \in \{0,1\}: i=1,\ldots,K-1\}$ is performed. In that context, the non-penalized PLS components are generalized linear combinations, of typte $t= \int_{T} X \left(t\right) w \left(t\right) dt,$ estimated by solving the following maximization problem
\begin{eqnarray}
\label{MAX_P1}
\{w,c\} &=& argmax \displaystyle Cov^{2}\left( \int_{T} X\left(t\right) w\left(t\right) dt, \sum_{i=1}^{K-1} c_i \tilde{Y}_i \right)\\
      & &\|w \| = 1 \nonumber \\
      & &\|c \| = 1 \nonumber
\end{eqnarray}
with $\| \cdot \|$ representing the usual norms in the spaces $L_2 [T]$ and $\mathbb{R}^{K-1}$ where the component weights belong to, respectively, and $c$ being the weight vector of $\tilde{Y}$.

From Proposition 1 in \cite{Preda2005}, it can be seen that the solution to this maximization problem is reached for $\pmb{w}$ and $\pmb{c},$ the eigenvectors
associated with the largest eigenvalue of $U_X$  and $U_{\tilde{Y}},$ respectively, defined by  $U_X = C_{X\tilde{Y}} \circ C_{\tilde{Y}X}$ and $U_{\tilde{Y}} = C_{\tilde{Y}X} \circ C_{X\tilde{Y}},$ with $C_{\tilde{Y}X}$ and $C_{X\tilde{Y}}$ being the
cross-covariance operators.
%

Then, the first PLS component, $t_{1} =  \int_{T} X \left( t\right) w_1 \left( t\right) dt,$ is estimated in terms of the associated weight function, $w_1 \left( t\right),$ which is solution to the following problem
$$C_{X\tilde{Y}} \circ C_{\tilde{Y}X} (w_{1}) = \lambda_{\max} w_{1}.$$

PLS regression is an iterative algorithm and the first PLS step is completed by ordinary linear regression of $X_{0}(t) =X(t)$ and $\tilde{Y}_{0}= \tilde{Y}$ on $t_{1},$ obtaining the residuals, $X_{1}(t)$ and $\tilde{Y}_{1},$ necessary to the following step
	\begin{eqnarray*}
             X_{1}(t) & = & X_{0}(t) - p_{1}(t) t_{1} \quad  t\in T \\
             \tilde{Y}_{1,i} & = & \tilde{Y}_{0,i} - c_{1,i} t_{1} \quad i=1,\dots,K-1,\\
         \end{eqnarray*}
where $p_{1}\left(t\right)= \left(\mathbb{E}(X_{0}(t)t_{1})/\mathbb{E}(t_{1}^{2})\right)$ and $c_{1}=\left(\mathbb{E}(Y_{0} t_{1})/\mathbb{E}(t_{1}^{2})\right).$

In general, let us consider the \emph{h}-th PLS component, $t_{h} = \int_{T}   X_{h-1} \left( t \right) w_h \left( t \right) dt,$ which is estimated by the following problem
$$C_{X\tilde{Y}}^{h-1} \circ C_{\tilde{Y}X}^{h-1} (w_{h}) = \lambda_{\max} w_{h},$$
where $C_{X\tilde{Y}}^{h-1}$ and $C_{\tilde{Y}X}^{h-1}$ are the cross covariance operators of $X_{h-1} (t)$ and $\tilde{Y}_{h-1}$, respectively. The \emph{h}-th PLS step is completed by ordinary linear regression of $X_{h-1}(t)$ and $\tilde{Y}_{h-1} $ on $t_{h},$ obtaining the residuals $X_{h}(t)$ and $\tilde{Y}_{h}$
 \begin{eqnarray*}
         X_{h}(t) & = & X_{h-1}(t) - p_{h}(t) t_{h} \quad  t\in
        T \\
        \tilde{Y}_{h,i} & = & \tilde{Y}_{h-1,i} - c_{h,i} t_{h} \quad i=1,\dots,K-1,\\
        \end{eqnarray*}
where $p_{h}\left(t\right)= \left(\mathbb{E}(X_{h-1}(t)t_{h})/\mathbb{E}(t_{h}^{2})\right)$ and $c_{h}=\left(\mathbb{E}(Y_{h-1} t_{h})/\mathbb{E}(t_{h}^{2})\right).$


\emph{A basis representation approach}


Let us consider now the basis representation of the functional predictor and the weight function, i.e.
$$
X\left(t\right) = \displaystyle \sum_{j=1}^{p} \alpha_{j} \phi_{j}\left(t\right) \quad \text{and} \quad
w\left(t\right) = \displaystyle \sum_{j=1}^{p} w_{j} \phi_{j}\left(t\right).
$$
As a consequence,  the maximization problem in Equation (\ref{MAX_P1}) can be written as follows:
$$
\{w,c\} = argmax \displaystyle \frac{\pmb{w}' \pmb{\Psi} \pmb{\Sigma}_{\alpha \tilde{Y}} \pmb{c} \pmb{c}' \pmb{\Sigma}_{\alpha \tilde{Y}}' \pmb{\Psi w}}{\pmb{w}'\pmb{\Psi w} + \pmb{c}'\pmb{c}},
$$
with $\pmb{w}=\left(w_{1},\ldots,w_{p}\right)'$ being the vector of basis coefficients of $w\left(t\right)$ and $\pmb{\Sigma}_{\alpha \tilde{Y}}=(\sigma_{ji})_{p\times K-1}$ being the cross-covariance matrix between $\pmb{\alpha}$ (the vector of basis coefficients of $X(t)$) and $\tilde{\pmb{Y}},$ where $\sigma_{ji}=E[\alpha_{j}\tilde{Y}_{i}],$ with
$\{j=1\ldots,p; \quad i=1,\ldots,K-1\}$.

The cross-covariance operators expressed in terms of the basis expansion of the functional predictor $X(t)$ are given by
 $$
\begin{array}{ll}
\mathcal{C}_{\tilde{Y}X}:& L^{2} \left(T\right) \rightarrow \mathbb{R}^{K-1}\\
                   & f= \displaystyle  \sum_{j=1}^p f_j \phi_{j}\left(t\right)  \rightarrow C_{\tilde{Y}X} (f) = \pmb{\Sigma}_{\alpha \tilde{Y}}' \pmb{\Psi} \pmb{ f}\\
\mathcal{C}_{X\tilde{Y}}:& \mathbb{R}^{K-1} \rightarrow L^{2} \left(T\right)\\
                   & \pmb{x} \rightarrow f\left(t\right)= \sum_{j=1}^{K-1} f_j \phi_{j}\left(t\right): \pmb{f} = \pmb{\Sigma}_{\alpha \tilde{Y}} \pmb{x},\\
\end{array}
$$
where $\pmb{f}= (f_1, \dots,f_p)'$ is the vector of basis coefficients of the function $f.$

Then, at the first PLS step, the weight function $w_{1}(t)$ is given by the eigenfunction associated with the largest eigenvalue of $U_X$ so that
\begin{equation}
\pmb{\Sigma}_{\alpha \tilde{Y}} \pmb{\Sigma}_{\alpha \tilde{Y}}' \pmb{\Psi} \pmb{w}_{1} = \lambda \pmb{w}_{1},
\label{diago1}
\end{equation}
and the weight vector $\pmb{c}_{1}$ is given by the eigenvector associated with the largest eigenvalue of $U_{\tilde{Y}}$ so that
$$
\pmb{\Sigma}_{\alpha \tilde{Y}}' \pmb{\Psi} \pmb{\Sigma}_{\alpha \tilde{Y}} \pmb{c}_{1} = \lambda \pmb{c}_{1}.
\label{diago2}
$$

Let us consider now the decomposition $\pmb{\Psi}= \left(\pmb{\Psi}^{1/2} \right) \left(\pmb{\Psi}^{1/2} \right)'$. Then,
$$\pmb{w}'_{1} \pmb{\Psi} \pmb{w}_{1}= \pmb{w}'_{1} \left(\pmb{\Psi}^{1/2} \right) \left(\pmb{\Psi}^{1/2} \right)' \pmb{w}_{1}= \tilde{\pmb{w}}'_{1}\tilde{\pmb{w}}_{1},$$ with $\tilde{\pmb{w}}_{1}=\left(\pmb{\Psi}^{1/2} \right)' \pmb{w}_{1}$ $(\pmb{w}_{1}=(\pmb{\Psi}^{-1/2})'\tilde{\pmb{w}}_{1}).$
This way, Equation (\ref{diago1}) can be expressed as follows
\begin{equation}
\left(\pmb{\Psi}^{1/2}\right)' \pmb{\Sigma}_{\alpha \tilde{Y}} \pmb{\Sigma}_{\alpha \tilde{Y}}' \pmb{\Psi}^{1/2} \tilde{\pmb{w}_{1}} = \lambda \tilde{\pmb{w}_{1}}.
\label{diago3}
\end{equation}

At the \emph{h}-th PLS step, the \emph{h}-th PLS component is estimated in terms of the associated weight function by solving to the following problem
\begin{equation}
\left(\pmb{\Psi}^{1/2}\right)' \pmb{\Sigma}_{\alpha_{h-1} \tilde{Y}_{h-1}} \pmb{\Sigma}_{\alpha_{h-1} \tilde{Y}_{h-1}}' \pmb{\Psi}^{1/2} \tilde{\pmb{w}_{h}} = \lambda \tilde{\pmb{w}_{h}}.
\label{diago3b}
\end{equation}
where $\pmb{\Sigma}_{\alpha_{h-1} \tilde{Y}_{h-1}}$ is the cross-covariance matrix between $\pmb{\alpha}_{h-1}$ (the vector of basis coefficients of $X_{h-1}(t)$) and $\tilde{\pmb{Y}}_{h-1},$ and $\tilde{\pmb{w}}_{h}=\left(\pmb{\Psi}^{1/2} \right)' \pmb{w}_{h}$ $(\pmb{w}_{h}=(\pmb{\Psi}^{-1/2})'\tilde{\pmb{w}}_{h}).$

In general, by considering Equations (\ref{diago3}) and (\ref{diago3b}) and taking into account that $\left(\pmb{\Psi}^{1/2}\right)' \pmb{\Sigma}_{\alpha \tilde{Y}}$ is the cross-covariance matrix between $\tilde{\pmb{Y}}$ and $(\pmb{\Psi}^{1/2})' \pmb{\alpha}$, it can be concluded that FPLS with multiple response is equivalent to  ordinary PLS of $\tilde{\pmb{Y}}$ on the random vector $(\pmb{\Psi}^{1/2})' \pmb{\alpha}.$

\subsection{Penalized multi-class FPLS regression}
\label{pls-pen}

As the biomechanical data analyzed in this paper are not smooth enough, a penalized approach for multi-class functional partial least squares (FPLS) regression is proposed.

Let us consider now the roughness penalty function $$Pen_{d}(w) = \int_{T} [D^{d}(w)(t)]^{2}dt,$$
 with $d=2$, as a measure of roughness of $w$ \citep{Osullivan1986}. By considering the basis expansion of $w(t),$
 the penalty function can be written as $Pen_{d}(w)= \pmb{w}' \pmb{P}_{d} \pmb{w},$
where $\pmb{w}=\left(w_{1}, \ldots, w_{p} \right)'$ is the vector of basis coefficients of $w\left(t\right)$ and $\pmb{P}_{d}$ the matrix of the cross inner products of the \emph{d}-order derivatives of the basis functions. As alternative, working in any context where regression on B-splines is useful, it can be considered
$\pmb{P}_{d}=\left(\pmb{\triangle}^{d}\right)' \pmb{\triangle}^{d},$ with $\pmb{\triangle}^{d}$ the matrix of \emph{d}-order differences between the adjacent basis coefficients \citep{Eilers1996}.

In this section the PLS components are estimated by solving the following problem
\begin{eqnarray*}
\{w,c\} &=& argmax \displaystyle Cov^{2}\left( \int_{T} X\left(t\right) w\left(t\right) dt, \sum_{i=1}^{K-1} c_i \tilde{Y}_i \right)\\
      & &\|w \|_{\lambda} = 1 \nonumber \\
      & &\|c \| = 1 \nonumber
\end{eqnarray*}
with $\|\cdot \|$ being the norm in the space $\mathbb{R}^{K-1}$ and $\| \cdot \|_{\lambda}$ being the norm associated with an inner product defined by
$\langle \pmb{f}, \pmb{g}\rangle _{\lambda}= \langle \pmb{f}, \pmb{g}\rangle + \lambda (\pmb{f}' \pmb{P}_d \pmb{g}),$
with $\pmb{f}=\left(f_{1}, \ldots, f_{p} \right)'$ and $\pmb{g}=\left(g_{1}, \ldots, g_{p} \right)'$ being  the vectors of basis coefficients of $f\left(t\right)$ and $g\left(t\right),$ respectively.

Then, the \emph{h}-th PLS component ($h > 1$), $t_{h} = \int_{T}   X_{h-1} \left( t \right) w_h \left( t \right) dt,$ is estimated by solving the problem
\begin{eqnarray}
\label{MAX_Pen1}
\{w,c\} &=& argmax \displaystyle Cov^{2}\left( \int_{T} X_{h-1}\left(t\right) w\left(t\right) dt, \sum_{i=1}^{K-1} c_i \tilde{Y}_{h-1,i} \right)\\
      & &\|w \|_{\lambda} = 1 \nonumber \\
      & &\|c \| = 1 \nonumber
\end{eqnarray}

By assuming the basis representation of the functional predictor and the weight function, the problem in (\ref{MAX_Pen1})
can be written as follows
\begin{equation}
\{w,c\} = argmax \displaystyle \frac{\pmb{w}' \pmb{\Psi} \pmb{\Sigma}_{\alpha_{h-1} \tilde{Y}_{h-1}} \pmb{c} \pmb{c}' \pmb{\Sigma}_{\alpha_{h-1} \tilde{Y}_{h-1}}' \pmb{\Psi w}}{\pmb{w}' (\pmb{\Psi} + \lambda \pmb{P}_d) \pmb{w}  + \pmb{c}'\pmb{c}},
\label{MAX_Pen2}
\end{equation}
with $\pmb{w}=\left(w_{1},\ldots,w_{p}\right)'$ being the vector of basis coefficients of $w\left(t\right),$ $\pmb{\Sigma}_{\alpha_{h-1} \tilde{Y}_{h-1}}=(\sigma_{ji})_{p\times K-1}$ being the cross-covariance matrix between $\pmb{\alpha_{h-1}}$ and $\tilde{\pmb{Y}}_{h-1},$ where $\sigma_{ji}=E[\alpha_{h-1,j}\tilde{Y}_{i}],$ with $\{j=1\ldots,p; \quad i=1,\ldots,K-1\},$ and $\lambda$ and $\pmb{P}_d$ the smoothing parameter and the penalty matrix, respectively.

Assuming the decomposition $\pmb{LL}'=\pmb{\Psi} + \lambda \pmb{P}_{d}$ and defining $\tilde{\pmb{w}}=\pmb{L}'\pmb{w}$ $(\pmb{w}=(\pmb{L}^{-1})'\tilde{\pmb{w}})$ the problem in (\ref{MAX_Pen2})
can be expressed as follows
$$
\{w,c\} = argmax \displaystyle \frac{\tilde{\pmb{w}}' \pmb{L}^{-1}\pmb{\Psi} \pmb{\Sigma}_{\alpha \tilde{Y}} \pmb{c} \pmb{c}' \pmb{\Sigma}_{\alpha \tilde{Y}}' \pmb{\Psi} (\pmb{L}^{-1})' \tilde{\pmb{w}}}{\tilde{\pmb{w}}' \tilde{\pmb{w}} +  \pmb{c}'\pmb{c}}.
$$

By analogy with the non-penalized approach, the associated eigenproblem is
\begin{equation}
\pmb{L}^{-1}\pmb{\Psi} \pmb{\Sigma}_{\alpha \tilde{Y}} \pmb{\Psi}' (\pmb{L}^{-1})' \tilde{\pmb{w}}=\lambda \tilde{\pmb{w}}, \quad \tilde{\pmb{w}}\in \mathbb{R}^{p}.
\label{eigenp_Pen}
\end{equation}
In general, the \emph{h}-th PLS component, $t_{h}=\int_{T}X(t)w_{h}(t)dt,$ is defined by $\pmb{w}_{h}=(\pmb{L}^{-1})'\tilde{\pmb{w}}_{h},$ which is the vector of basis coefficients of $w_{h}(t),$ with $\tilde{\pmb{w}}_{h}$ being the eigenvector related to the largest eigenvalue of the problem in (\ref{eigenp_Pen}).

Finally, taking into account that $\pmb{L}^{-1}\pmb{\Psi} \pmb{\Sigma}_{\alpha \tilde{Y}}$ is the cross-covariance matrix between
$\tilde{\pmb{Y}}$ and $\pmb{L}^{-1}\pmb{\Psi} \pmb{\alpha},$ the penalized approach is reduced to a
classical PLS of $\tilde{\pmb{Y}}$ on the random vector $\pmb{L}^{-1}\pmb{\Psi} \pmb{\alpha}.$

\subsection{The problem of repeated measures}
\label{ML-FPLS}

In previous sections, penalized and non-penalized approaches for the multi-class functional partial least squares regression (FPLS) have been proposed.
However, these approaches are based on ordinary PLS, and then the between-subject and the within-subject variations
are not studied by separate.

As solution to the problem of repeated measures, in this section a multi-class approach for penalized and non-penalized FPLS regression based on the split-up variation is proposed. In the same spirit as in \citet{multilevel1,multilevel2,multilevel3}, this approach first decomposes the variability in the data matrix and then applies the multi-class PLS regression on the within-subject variation matrix.

For a more detailed explanation of the multi-class approach, let us consider $\{x_{i}\left(t\right): t\in T, i=1,\ldots, n\}$ a sample of the
functional variable $X(t)$ and $\{y_{1},y_{2},\ldots,y_{n}\}$ a random sample of $Y$ associated with it, with $y_{i}=\{1,2,\ldots,K\}$.
Henceforth, let $\tilde{Y}$ denote a $N\times (K-1)$ matrix with zeros and ones, comprising the values associated with the dummy variables obtained from $Y$.

Then, the sample estimation of the multi-class FPLS described in sections \ref{pls} (non-penalized) and \ref{pls-pen} (penalized) is as follows:
\begin{itemize}
\item Non-penalized multi-class FPLS: multi-class PLS of $\tilde{\pmb{Y}}$ on the matrix $\pmb{X}^{1}_{(N\times p)}=\pmb{A}\pmb{\Psi}^{1/2}.$
\item Penalized multi-class FPLS: multi-class PLS of $\tilde{\pmb{Y}}$ on the matrix $\pmb{X}^{2}_{(N\times p)}=\pmb{A}\pmb{\Psi}(\pmb{L}^{-1})',$ so that $\pmb{LL}'=\pmb{\Psi} + \lambda \pmb{P}_{d},$
\end{itemize}
with $\pmb{A}$ being the matrix of basis coefficients of the functional variable $X,$ and $\pmb{\Psi}$ and $\pmb{P}_{d}$ as in previous sections.

In practice functional data are observed in a finite set of points. Because of this, matrix $A$ must be estimated. By assuming that the underline process to biomechanical data is smooth, B-spline basis functions are considered, and the basis coefficients are estimated by the least squares criterion. This approach is called regression splines. Additionally, for data observed with noise, a penalized estimation of the basis coefficients in terms of P-splines can be considered \citep[see][for more details]{AguileraPspl2013,Eilers1996}.

\noindent  \emph{Split-up variation}

In general, let us consider a data matrix $\pmb{X}_{N\times p}$ containing the information related to $n$ subjects and each of the $K$ possible stimulus ($N=n\times K$), with $p$ being the number of basis functions (or variables in the multivariate setting). In that sense, let $x_{i k}^{j}$ denote the value of the \emph{j}-th column of $X$ for the \emph{i}-th subject under the \emph{k}-th stimulus.

Following the mixed-effect models philosophy, matrix $X$ can be decomposed into an offset term, a between-subject and a within-subject part so that
$$\pmb{X}= \pmb{X}_{offset} + \pmb{X}_{between-subject} + \pmb{X}_{within-subject}.$$

But mixed models relies on assumptions such as Gaussian distribution of random effects. As an alternative, a split-up variation approach, which does not require the above-mentioned assumptions, is considered in this paper. Exactly, the offset, between-subject and within-subject variation matrices ($\pmb{X}_{o},$ $\pmb{X}_{b}$ and $\pmb{X}_{w},$ respectively) can be obtained as follows: $\pmb{X}_{o}=\pmb{1}_{(N\times 1)} \pmb{x}'_{\cdot \cdot (1\times p)},$ where $\pmb{x}_{\cdot \cdot}=(x_{\cdot \cdot}^{1}, \ldots, x_{\cdot \cdot}^{p})'$ and $x_{\cdot \cdot}^{j}=\frac{1}{N}\sum_{k=1}^{K} \sum_{i=1}^{n} x_{i k}^{j}.$ $\pmb{X}_b$ is obtained by concatenating the matrices $\pmb{1}_{(K\times 1)} \pmb{x}'_{b i (1\times p)}$ for each subject into $\pmb{X}_b,$ where
$\pmb{x}_{b i}=(x_{i \cdot}^{1} - x_{\cdot \cdot}^{1}, \ldots, x_{i \cdot}^{p} - x_{\cdot \cdot}^{p})'$ and
 $x_{i \cdot}^{j}=\frac{1}{K}\sum_{k=1}^{K}  x_{i k}^{j}.$ $\pmb{X}_w=\pmb{X} - \pmb{X}_{i \cdot},$ with $\pmb{X}_{i \cdot}$  obtained by concatenating the matrices
$\pmb{1}_{(K\times 1)} \pmb{x}'_{i \cdot (1\times p)}$ for each subject into $\pmb{X}_{i \cdot},$ with $\pmb{x}_{i \cdot}=(x_{i \cdot}^{1}, \ldots, x_{i \cdot}^{p})'.$

\noindent  \emph{Multi-class approach for FPLS}

Once the split-up variation has been carried out, penalized and non-penalized multi-class FPLS regression is performed on the within-subject variation matrix.
In that sense, the penalized and non-penalized multi-class approaches for FPLS regression are given by
\begin{itemize}
\item Non-penalized multi-class FPLS: multi-class  PLS of $\tilde{\pmb{Y}}$ on the matrix $\pmb{X}^{1}_{w}$
\item Penalized multi-class FPLS: multi-class  PLS of $\tilde{\pmb{Y}}$ on the matrix $\pmb{X}^{2}_{w},$
\end{itemize}
where $\pmb{X}^{1}_{w}$ and $\pmb{X}^{2}_{w}$ are the within-subject variation matrices from $\pmb{X}^{1}$ and $\pmb{X}^{2},$ respectively.

\subsection{Functional LDA on a reduced set of functional PLS components}

Let $\pmb{T}$ be a $n\times q$ matrix comprising the columns of the first $q$
PLS component scores (by variability order). By considering the sample estimation
of the penalized and non-penalized multi-class approaches for functional partial least squares (FPLS) regression proposed in Section \ref{ML-FPLS},
the sample estimation of $\pmb{T}$ is as follows
\begin{itemize}
    \item Non-penalized multi-class FPLS: $\pmb{T}=\pmb{A \Psi}^{1/2}\pmb{V}$
    \item Penalized multi-class FPLS: $\pmb{T}=\pmb{A \Psi} (\pmb{L}^{-1})'\pmb{V},$
\end{itemize}
with $\pmb{V}$ being the matrix comprising the columns of the eigenvectors $\tilde{\pmb{w}}_{1}, \ldots, \tilde{\pmb{w}}_{q}$ associated with the $t_{1},\ldots, t_{q}$ PLS components.

Once the PLS components have been estimated,
linear discriminant analysis (LDA) of the original sample values of $Y$ $(y_{1},\ldots,y_{n})$ on the matrix $\pmb{T}$ is carried out.
Then, linear combinations such as $\Phi_{i}(X)=\alpha_{i} + \pmb{T}\pmb{\beta}_{PLS}^{i},$ $\{i=1,\ldots,K-1\}$ are estimated,
so that the between-class variance is maximized with respect to the within-class variance, with
$\pmb{\beta}_{PLS}^{i}=(\beta_{PLS_{1}}^{i},\ldots,\beta_{PLS_{q}}^{i})'$ being the vector of discriminant coefficients.

But the aim is to estimate the discriminant functions $\beta^{i}(t),$ $i=1,\ldots,K-1$ included in the linear combinations
given in equation (\ref{flda}). To this end, let us consider the basis representation of the sample curves and the discriminant functions $(\beta^{i}(t)=\sum_{j=1}^{p}\beta_{j}^{i}\phi_{j}(t))$
and, as a consequence, equation (\ref{flda}) can be rewritten as
$$ \Phi_{i}(X)=\alpha_{i} + \int_{0}^{T}X(t)\beta^{i}(t)dt = \alpha_{i} + \pmb{A\Psi \beta}^{i}, $$
with $\pmb{\beta}^{i}$ being the vector of basis coefficients of the \emph{i}-th discriminant function.

Finally, taking into account the sample estimation of $\pmb{T},$ we can conclude that
\begin{itemize}
	\item Non-penalized multi-class FPLS: $\pmb{\beta}^{i} = (\pmb{\Psi}^{-1/2})'\pmb{V}\pmb{\beta}_{PLS}^{i}, \; i=1,\ldots,K-1$
	\item Penalized multi-class FPLS: $\pmb{\beta}^{i} = (\pmb{L}^{-1})'\pmb{V}\pmb{\beta}_{PLS}^{i}, \quad i=1,\ldots,K-1.$
\end{itemize}

Once the random vector of discriminant functions $\beta(t)=(\beta^{1}(t), \ldots, \beta^{K-1}(t))'$ is obtained, the classification is then performed in the
transformed space based on some distance metric, such as Euclidean distance.

Then given a new sample observation $x_{0}(t),$ it is classified to
\begin{equation}
\begin{aligned}
& \underset{k}{\text{argmin}}
& & d\left(\int_{T} x_{0}(t) \beta(t), \int_{T} \overline{x}^{k}(t) \beta(t)\right),\\
\end{aligned}
\label{euclidean_dist}
\end{equation}
where $\overline{x}^{k}(t) = \frac{1}{m_k} \sum_{x(t)\in C_k} x(t)$ is the mean curve related to the \emph{k}-th class and $m_k$ the number of sample observations in the \emph{k}-th class, with $k=1,\ldots,K.$

By assuming the basis representations of $x_{0}(t),$ $\beta(t)$ and $\overline{x}^{k}(t)$
\begin{equation*}
\begin{aligned}
& x_{0}(t)=\sum_{j=1}^{p} a_{0_j}\phi_{j}(t),
& & \beta^i(t)= \sum_{j=1}^{p} \beta^{i}_{j}\phi_{j}(t),
& & & \overline{x}^{k}(t)=\sum_{j=1}^{p} \overline{a}^{k}_{j}\phi_{j}(t),
\end{aligned}
\end{equation*}
problem (\ref{euclidean_dist}) can be rewritten as follows
\begin{equation*}
\begin{aligned}
& \underset{k}{\text{argmin}}
& & d\left(\pmb{a}_{0}' \pmb{\Psi} \pmb{B}, \pmb{\overline{a}}^{k'} \pmb{\Psi} \pmb{B} \right),\\
\end{aligned}
\end{equation*}
with $\pmb{a}_{0_{p\times 1}}$ being the vector of basis coefficients of $x_{0}(t),$ $\pmb{\overline{a}}^{k}_{p\times 1}$ being the vector of basis coefficients of $\overline{x}^{k}(t),$ $\pmb{\Psi}_{p\times p}$ being the inner product matrix, $\pmb{B}_{p\times (K-1)}$ the
matrix with columns the vectors of basis coefficients of the $K-1$ discriminant functions (i.e., $\pmb{\beta}^{1}, \ldots, \pmb{\beta}^{K-1}$). Finally,
the Euclidean distance between two vectors $\pmb{V}^{1}=\pmb{a}_{0'} \pmb{\Psi} \pmb{B}$ and $\pmb{V}^{2}=\pmb{\overline{a}}^{k'} \pmb{\Psi} \pmb{B}$ is computed as $d(\pmb{V}^{1}, \pmb{V}^{2})= \sqrt{\sum_{i=1}^{K-1} (\pmb{V}^{1}_{i}- \pmb{V}^{2}_{i})^2}.$

The smoothing parameter $\lambda$ and the number of PLS components used in the LDA approach based on Penalized multi-class FPLS have been selected by leave-one-out cross-validation on a training sample, selecting the combination of both that minimizes the misclassification rate obtained from LDA on the PLS components. In the non-penalized approach, only the number of PLS components is required, and it is chosen by leave-one-out cross-validation on a training sample, minimizing the misclassification rate obtained from  functional LDA.

All the analysis in Sections \ref{sim} and \ref{results} were carried out in a computer with an Intel Core i5 processor at 3 GHz with 8Gb of RAM, running R-project version 3.5.1 \citep{Rproject}. The two multi-class functional PLS approaches were reduced to multi-class PLS regression (in the multivariate context), and then they were implemented in R using the function \emph{spls} available in the \emph{mixOmics} R package developed by \citet{mixomicsR}.

Authors are preparing an R package that, among other things, includes the code used in the manuscript. Then, readers wishing to use the software, it will be available as soon as possible at the R repository website. However, you can contact the corresponding author if you need more information regarding the code.

\section{Simulation study}
\label{sim}

In order to test the performance of the proposed methods, a simulation study has been carried out. Inspired by \cite{Cuevas2004}, let us consider $120$ sample curves, $40$  for each of the 3 classes, according to the following process
$$X_{i}^{k}(t) = m_{k}(t) + \epsilon_{i}^{k}(t), \; t \in [0,1], \quad i=1,\ldots,40, \quad k=1,2,3,$$
where $m_{i}(t) = t^{(k/5)}(1-t)^{(6-k/5)}.$ The process $X_{i}^{k}(t)$ was generated in discretized version $X_{i}^{k}(t_{j}),$ for $j=1,\ldots,101$, with $t_{1},\ldots,t_{101}$ being equidistant points in $[0,1].$ Finally, $\epsilon_{i}^{k}(t_{j})$ are i.i.d. random variables $N(0, \sigma_{\epsilon}=0.2).$

But our work is focused on the presence of repeated measurements in the $k$ classes, and then an additional term  must be considered in order to represent the subject effect. In that sense, and inspired by \cite{Durban2005}, a specific-subject term has been included in the above process so that
$$X_{i}^{k}(t) = m_{k}(t) + a_{i}\sin(t\pi)+  \epsilon_{i}^{k}(t), \; t \in [0,1], \quad i=1,\ldots,40, \quad k=1,2,3,$$
with $a_{i}\sim N(\mu_{i}, \sigma_{s}=0.02)$ and $\mu_{i}\sim Uniform(0,0.05).$ The $\sin(t\pi)$ function has been considered in order to simulate curves taking the same value at the beginning and at the end of the observation interval (as in the gait data study). Of the 120 simulated sample curves, 90 (30 for each class) were considered as training sample and the remaining 30 (10 for each class) as test sample.

\begin{center}
\begin{figure}[ht]
\includegraphics[width=.80 \textwidth, height =.40 \textheight]{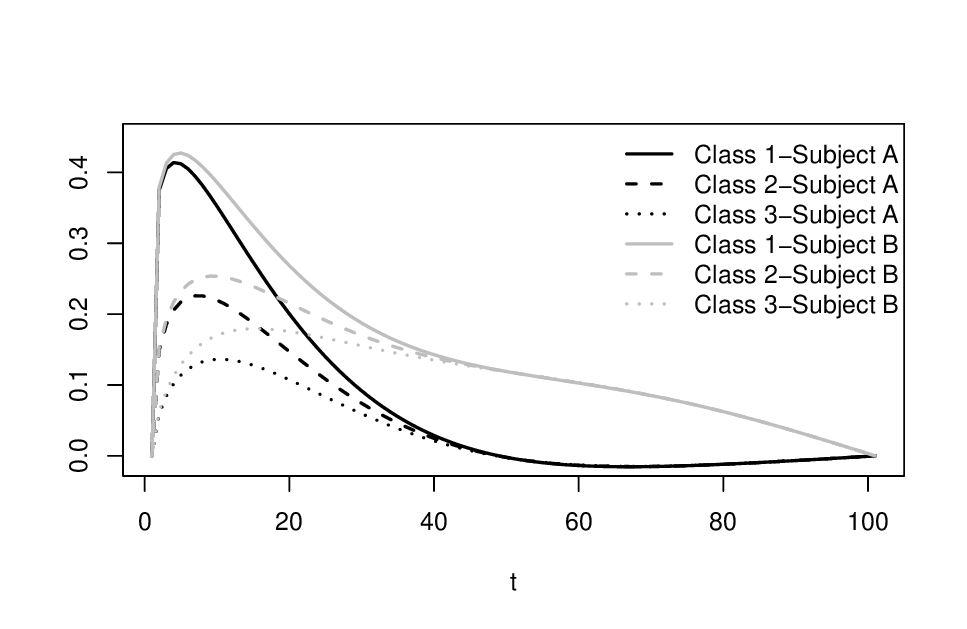}\\
\caption{Simulation study. Sample paths related to classes 1, 2 and 3 (solid, dashed and dotted line, respectively) for two subjects A and B, black and grey lines, respectively.}
\label{Fig:sim1}
\end{figure}
\end{center}

The sample curves from two different subjects, and for each of the three classes, are shown in Figure \ref{Fig:sim1}. As we can see, there is a clear dependency structure between the three observations of the same subject, and the between subjects variation is not only given by a magnitude effect, but also by a shape variation. This is a realistic simulation of the performance of kinematic data since it makes no sense to assume that the curves of the different subjects are parallel.

\begin{center}
\begin{figure}[ht]
\begin{center}
\begin{tabular}{cc}
\includegraphics[width=.47 \textwidth, height =.30 \textheight]{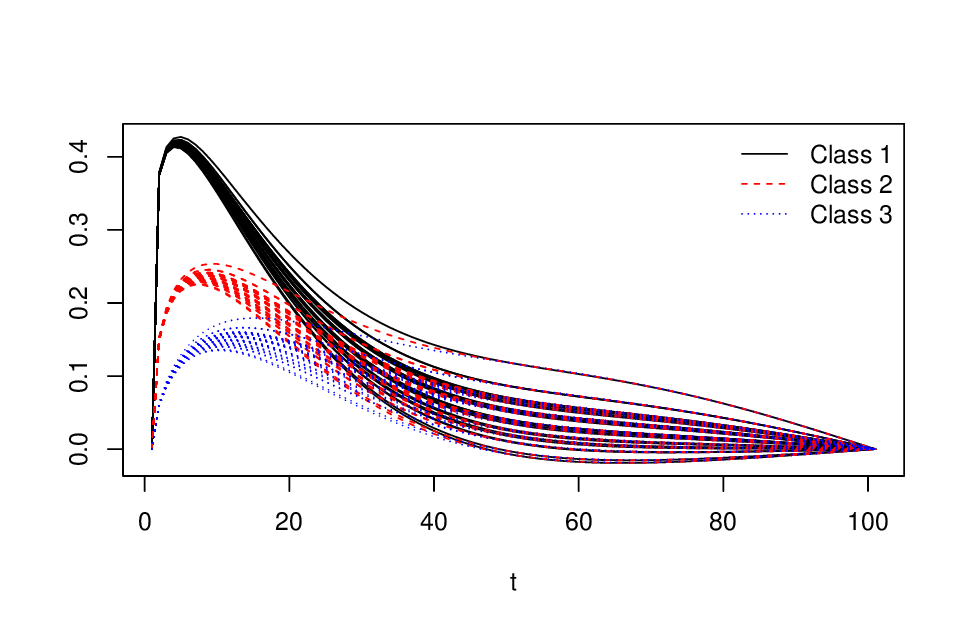}&
\includegraphics[width=.47 \textwidth, height =.30 \textheight]{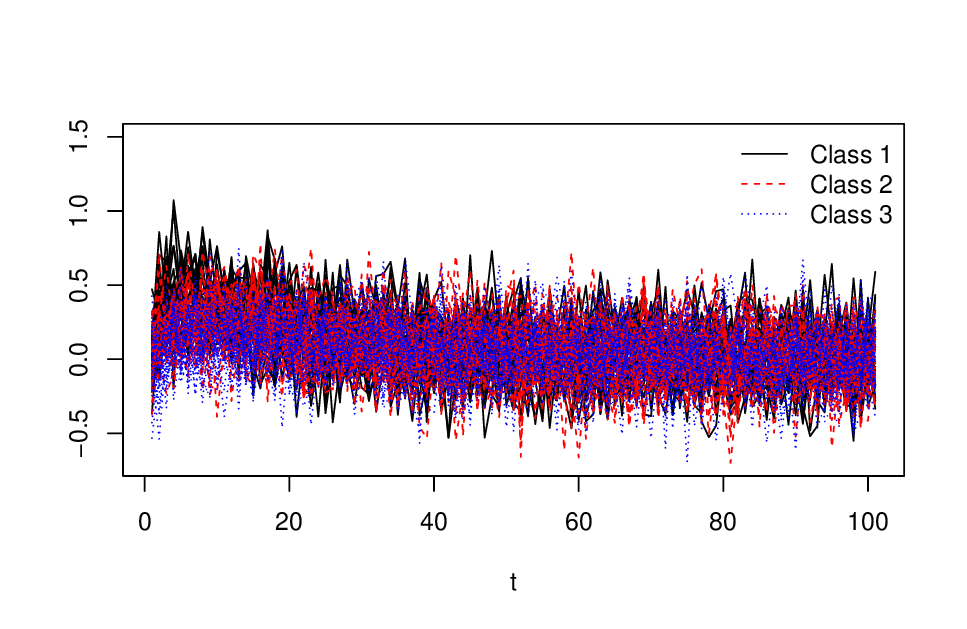}\\
\end{tabular}
\end{center}
\caption{Simulation study. Smooth and noisy sample paths (left and right panel, respectively) related to classes 1, 2 and 3 (solid line, dashed line and dotted line, respectively) for 30 subjects.}
\label{Fig:sim2}
\end{figure}
\end{center}

In Figure \ref{Fig:sim2} all the sample curves, with and without noise, are displayed highlighting the three classes by different line types. In order to fit the regression splines, a cubic B-spline basis defined on 15 basis knots has been considered.

\begin{center}
\begin{figure}[ht]
\begin{center}
\begin{tabular}{cc}
$CCR_{training}$ & $CCR_{test}$\\
\includegraphics[width=.45 \textwidth, height =.30 \textheight]{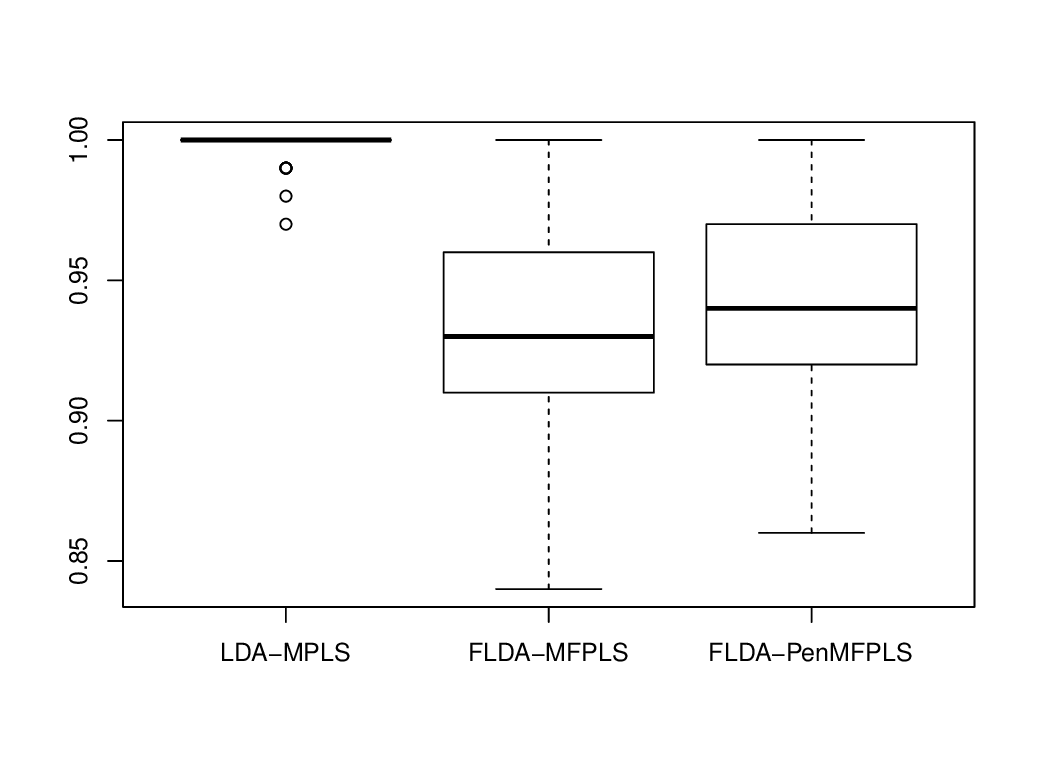}&
\includegraphics[width=.45 \textwidth, height =.30 \textheight]{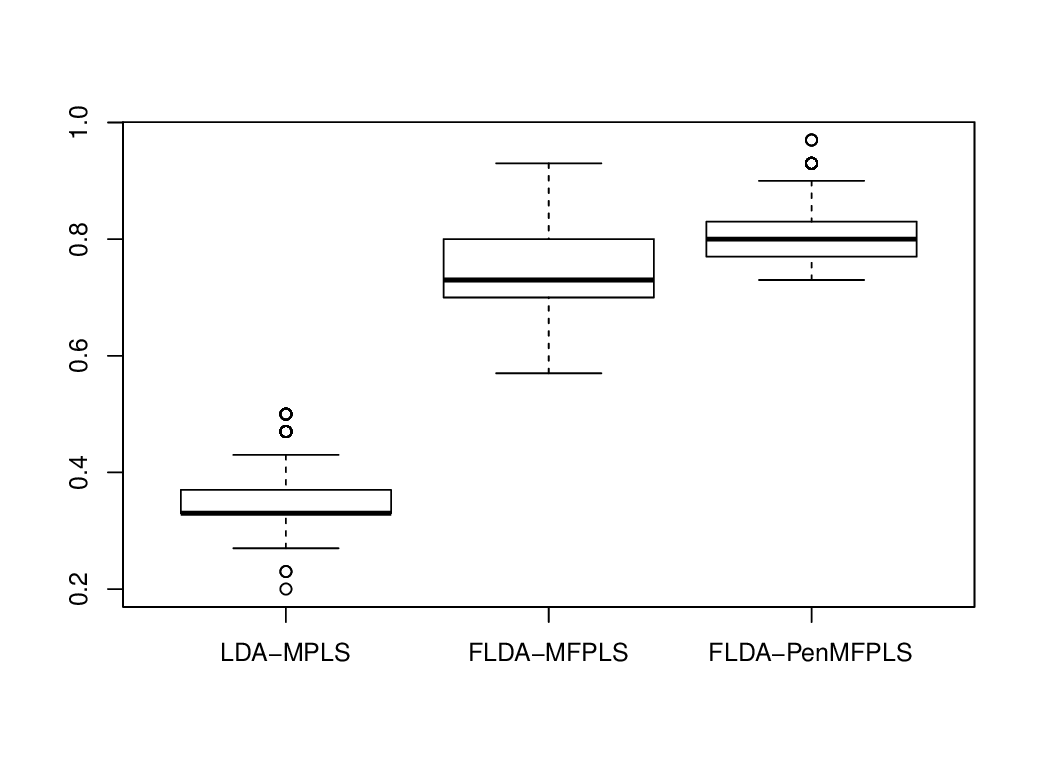}\\
\end{tabular}
$\#PLS cp's$ \\
\includegraphics[width=.45 \textwidth, height =.30 \textheight]{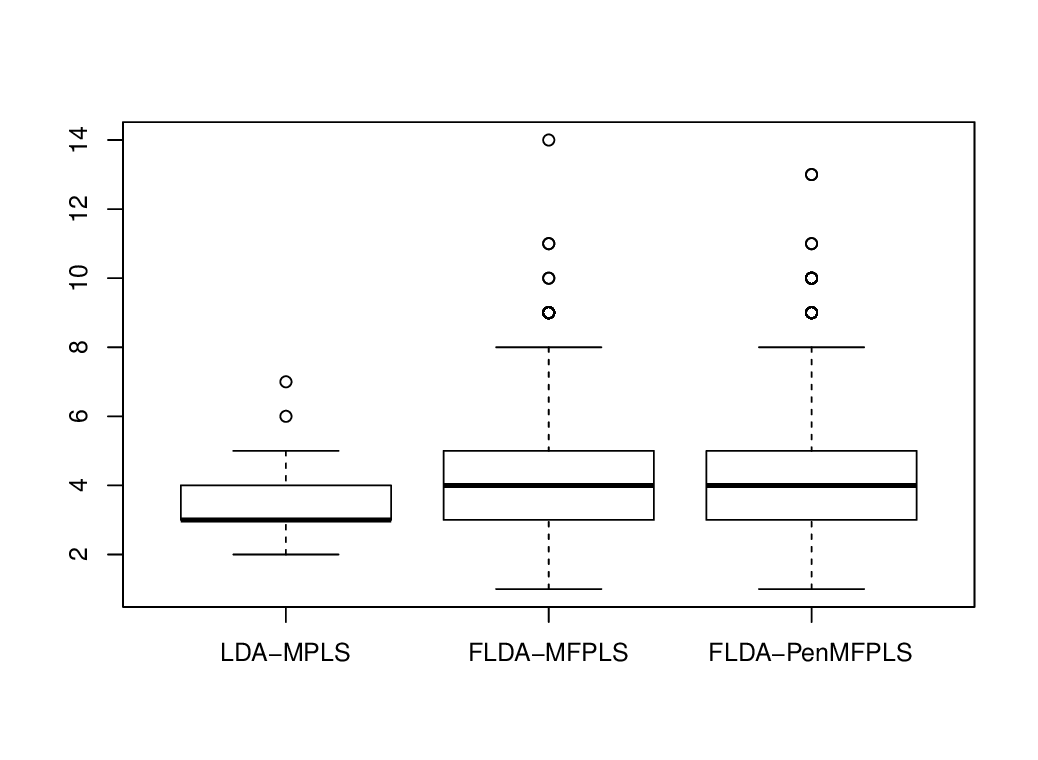}\\
\end{center}
\caption{Simulation study. Box plots showing to the number of PLS components used in the LDA (bottom panel), the correct classification rates from the cross-validation on the training sample and from the classification of the test sample (top-left and top-right panels, respectively). Experiments run 500 times.}
\label{Fig:sim3}
\end{figure}
\end{center}

The simulation scheme was run 500 times and the main results are summarized in Figure \ref{Fig:sim3}. As we can see, the two functional approaches tend to select a larger number of PLS components than the multivariate version, as covariates in the LDA. But there is not a great difference between them. Regarding the correct classification rates (CCR), there is a clear overfitting in the multivariate approach, which provides CCR close to 1 in the cross-validation. Between the two proposed functional approaches, the one based on penalized multi-class functional PLS regression provides the best performance, with the highest CCR and the minimum variability.

\section{Case studies}
\label{results}

In this section the results from the analysis of two real kinematic data sets are summarized. In general, kinematic data from inertial sensors placed on human body is usually collected in the format of linear acceleration or angular rotation. Linear acceleration or angular rotation represent the trajectory of the human movement, which can be identified with a function in time and frequency domain. In that sense, data from kinematic studies can be considered as a functional data set, and then functional data techniques can be applied to analyze this type of data. In addition, it is important to remark the presence of repeated measures in kinematic studies. So, both the between-subject and within-subject variations must be taken into account.

\subsection{Human activity data}
\begin{center}
\begin{figure}[ht]
\begin{center}
\begin{tabular}{ccc}
\includegraphics[width=.31 \textwidth, height =.20 \textheight]{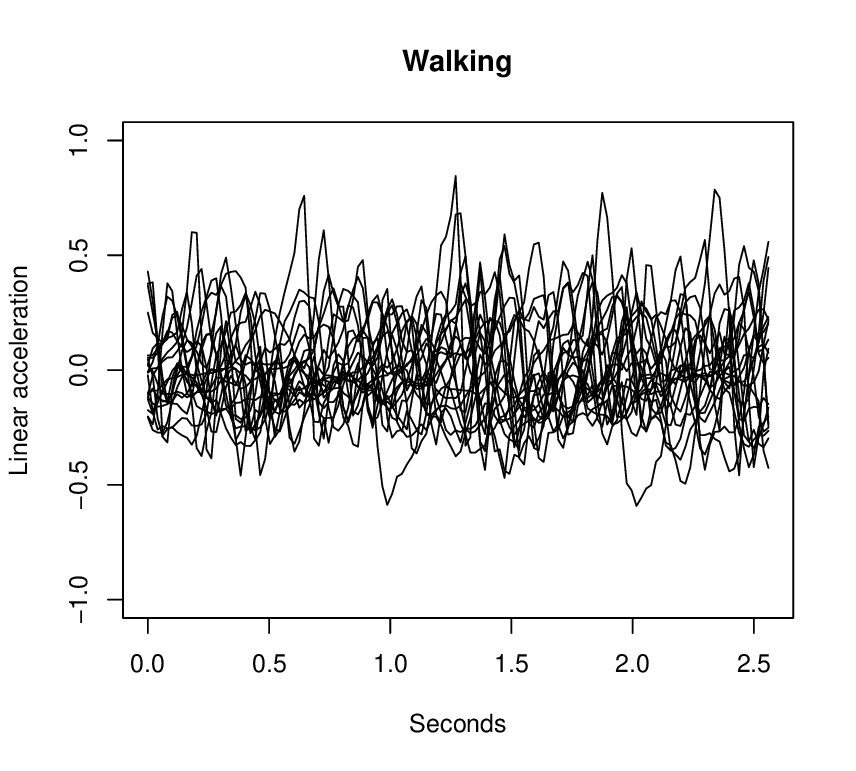}&
\includegraphics[width=.31 \textwidth, height =.20 \textheight]{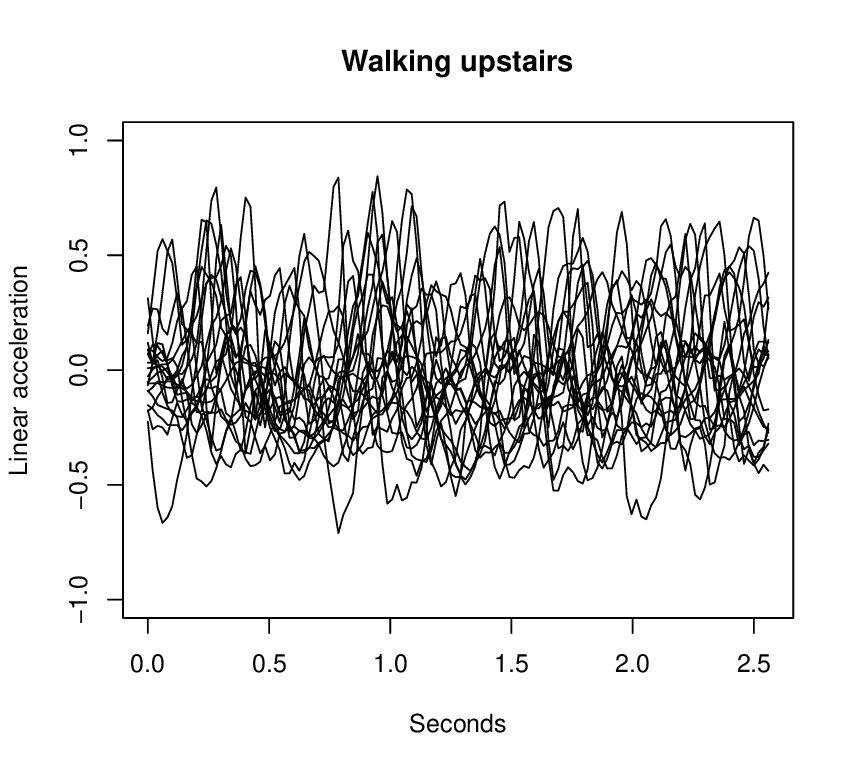}&
\includegraphics[width=.31 \textwidth, height =.20 \textheight]{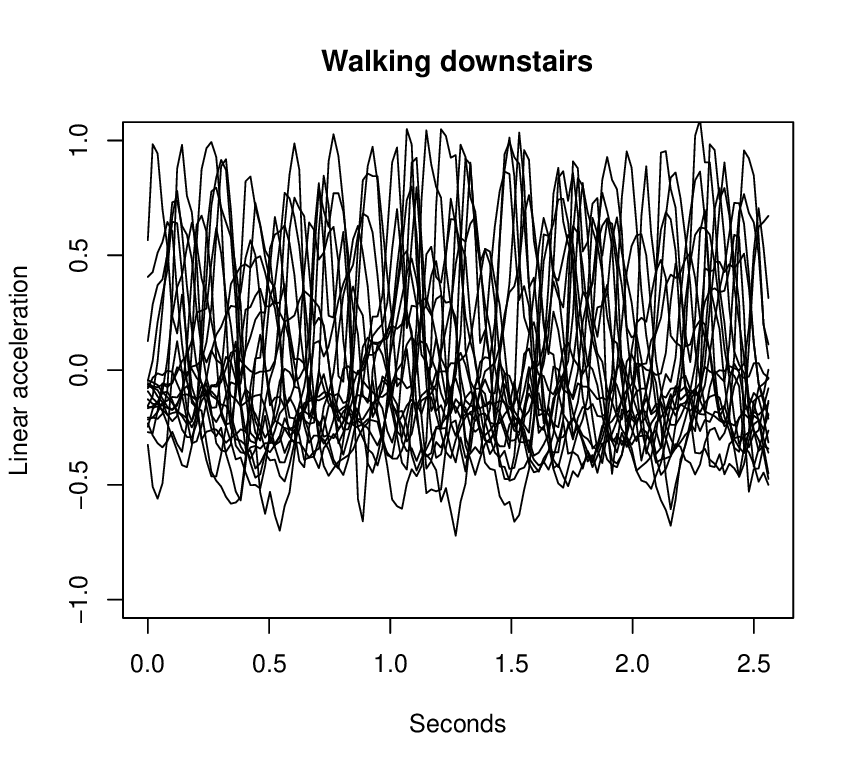}\\
\end{tabular}
\end{center}
\caption{Human activity data. Raw data. Sample paths displayed separately by stimulus: walking, walking upstairs and walking downstairs, from left to right, respectively.} \label{Fig:1}
\end{figure}
\end{center}

\begin{center}
\begin{figure}[ht]
\begin{center}
\begin{tabular}{cc}
\includegraphics[width=.45 \textwidth, height =.30 \textheight]{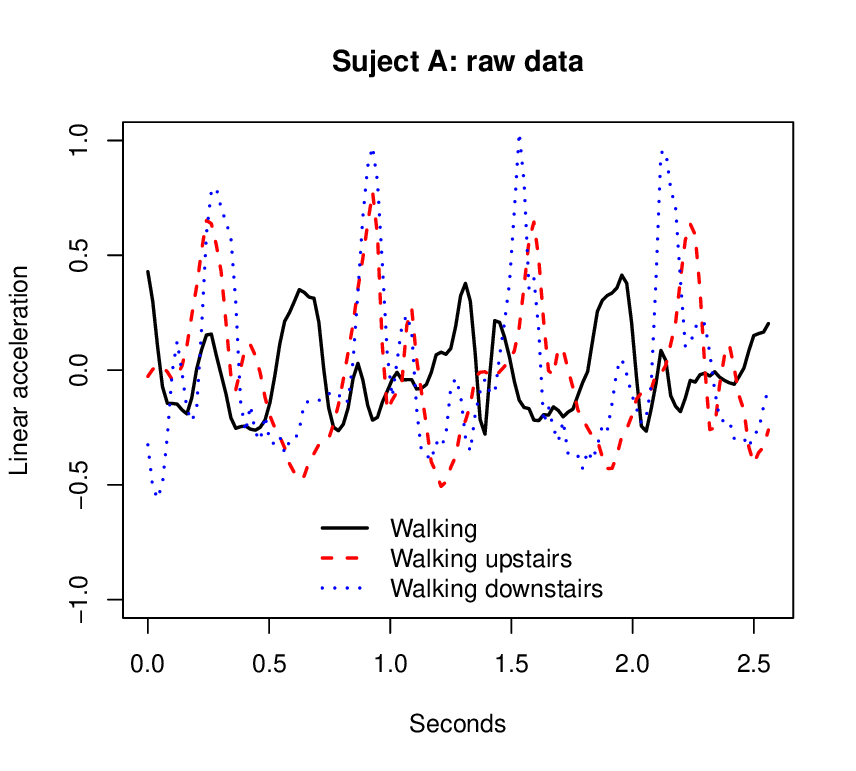}&
\includegraphics[width=.45 \textwidth, height =.30 \textheight]{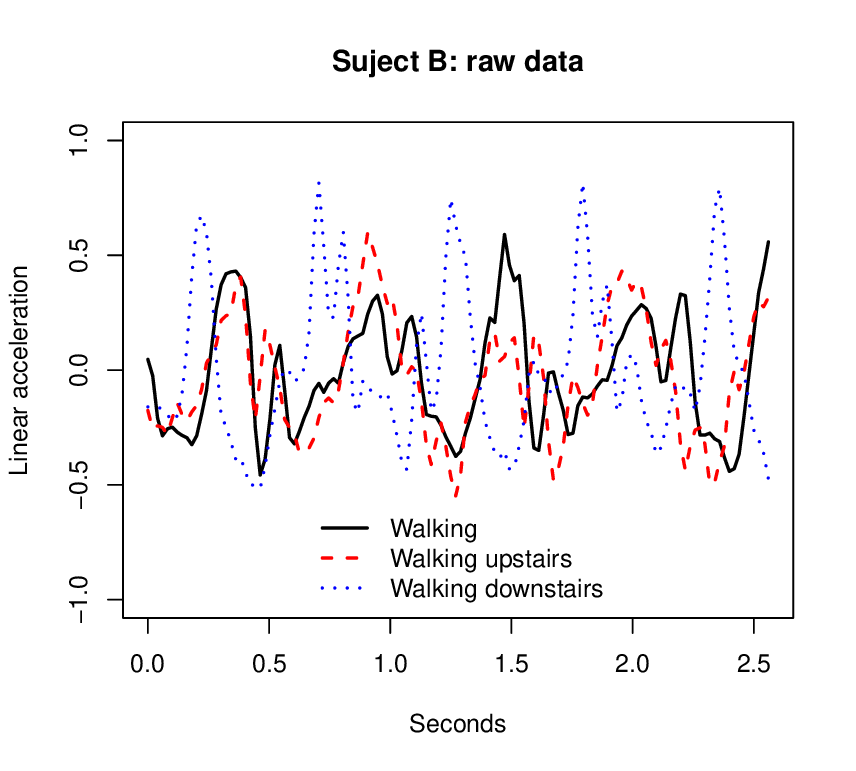}\\
\end{tabular}
\end{center}
\caption{Human activity data. Raw data. Sample paths related to walking (solid line), walking upstairs (dashed line) and walking downstairs (dotted line) for two subjects A and B, left and right panel, respectively.}
\label{Fig:1b}
\end{figure}
\end{center}

The human activity data set is part of a wider experiment, focused on the human activity recognition, carried out by \citet{HumanActivity}. In this paper let us consider the linear acceleration (meter per second squared), measured on axis X and recorded in 128 equidistant knots at the interval $[0, 2.56]$ seconds, and related to three different stimulus (walking, walking upstairs and walking downstairs). A total of  30 subjects participated in this experiment. Then, a total of 90 sample paths are considered (one sample path per stimulus and subject). Subject 23 was removed from the sample for being an outlier. The remaining sample observations were distributed into training and test sample, with observations (repeated measures) related to 20 and 9 subjects, respectively. The raw data (displayed by stimulus) are shown in Figure \ref{Fig:1}. In order to visualize the variability of the spectra between stimulus, in Figure \ref{Fig:1b} the spectra related to the three activities have been overlapped for two subjects A and B, left and right panel, respectively.

\begin{center}
\begin{figure}[ht]
\begin{center}
\begin{tabular}{ccc}
\includegraphics[width=.30 \textwidth, height =.20 \textheight]{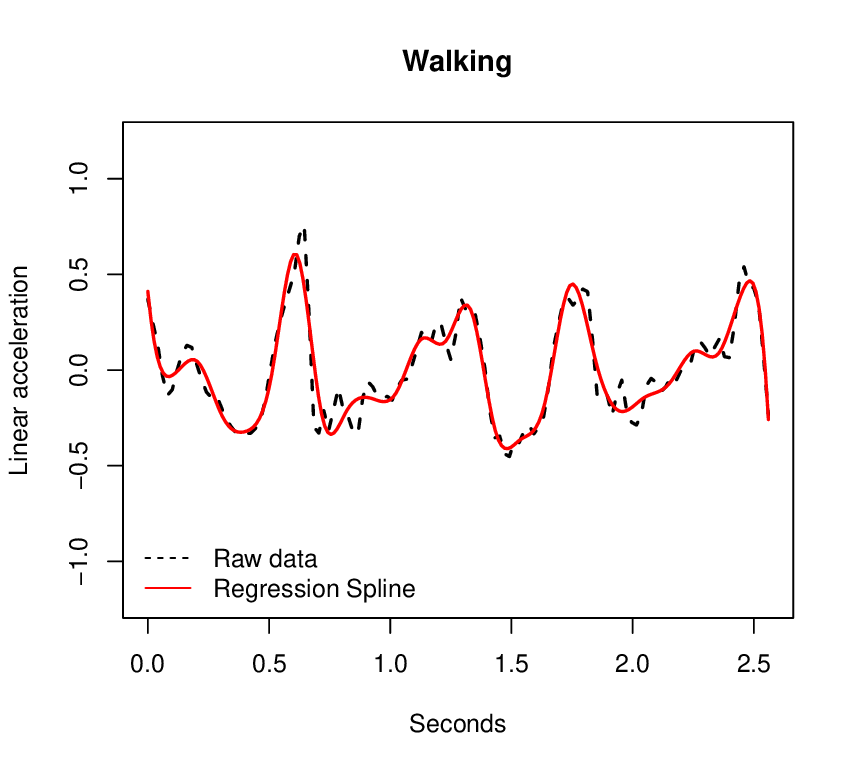}&
\includegraphics[width=.30 \textwidth, height =.20 \textheight]{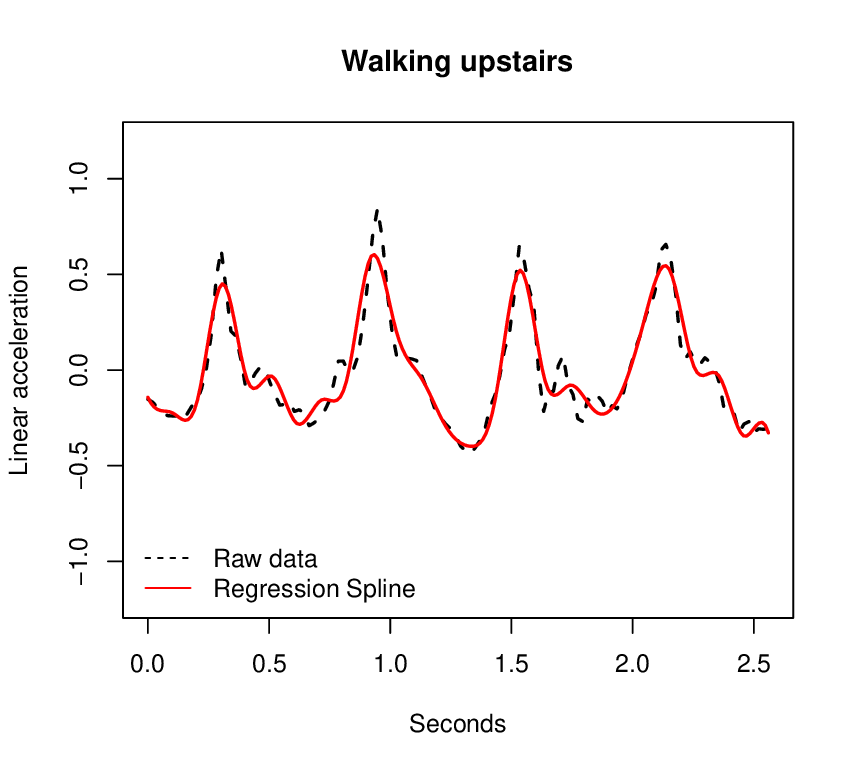}&
\includegraphics[width=.30 \textwidth, height =.20 \textheight]{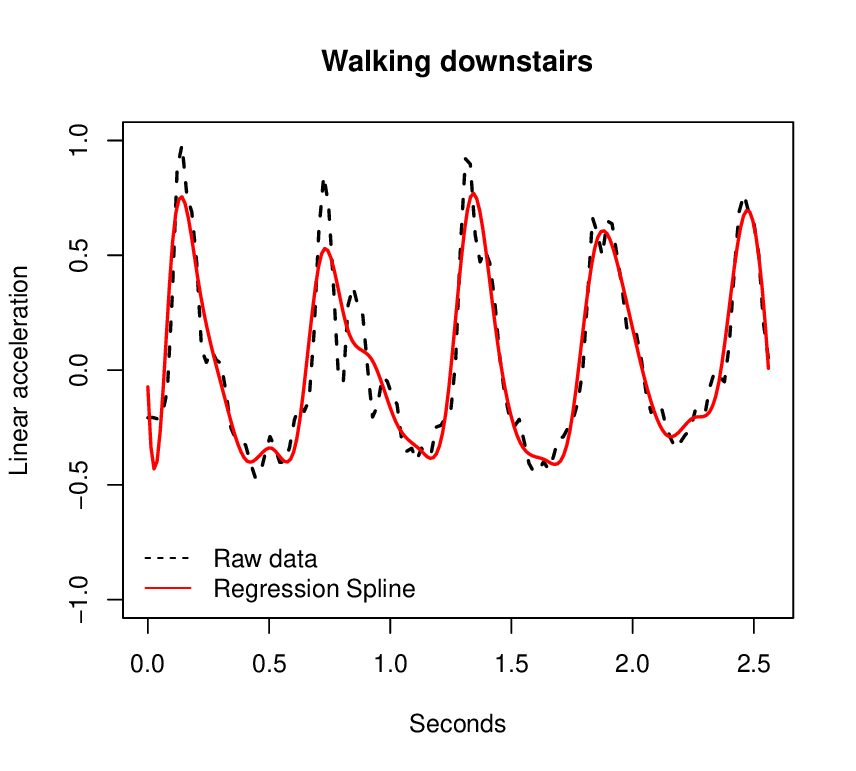}\\
\end{tabular}
\end{center}
\caption{Human activity data. A sample path for each stimulus (dashed line) together with the regression spline (solid line) estimated with a cubic B-splines basis defined on 25 equidistant knots.} \label{Fig:2}
\end{figure}
\end{center}

The first step in the analysis was to approximate the sample curves by mean of a basis representation and using a cubic B-spline basis defined on 25 equidistant knots.  In Figure \ref{Fig:2} the raw data together with the regression spline can be seen. Once the sample curves have been approximated, the aim is the human activity pattern recognition. To this end functional linear discriminant analysis based on a multi-class approach for functional PLS regression (FLDA-MFPLS), was carried out to classify the sample curves according to the stimulus which produced them. It could be assumed that the raw data, which were collected by a smartphone on the waist of the subjects, are affected by some error or noise. In that sense, and aiming to avoid a possible lack of smoothness in the estimation of the discriminant functions, the penalized version of FLDA-MFPLS (FLDA-PenMFPLS), proposed in section \ref{ML-FPLS}, was also considered.

\begin{center}
\begin{figure}[ht]
\begin{center}
\begin{tabular}{cc}
\includegraphics[width=.45 \textwidth, height =.28 \textheight]{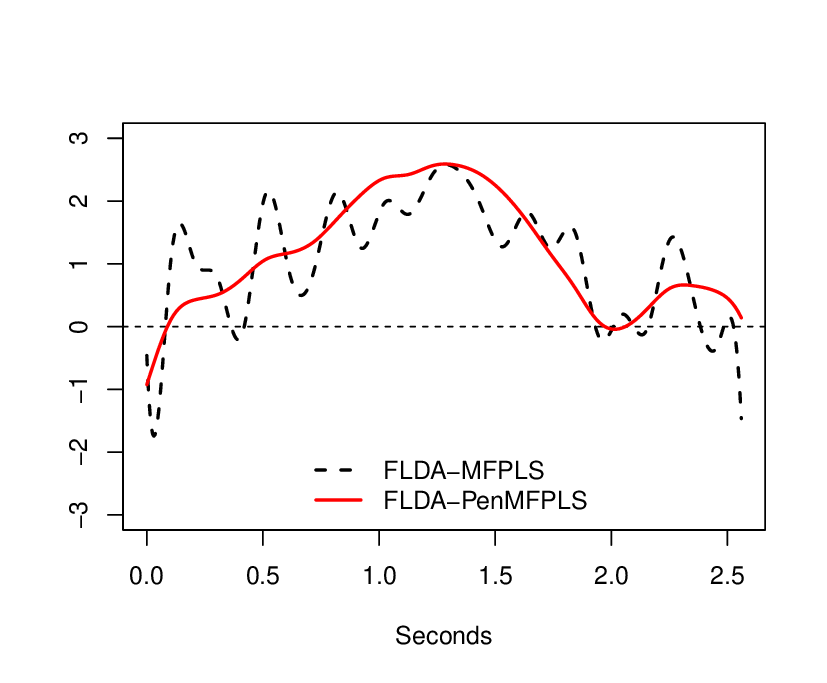}&
\includegraphics[width=.45 \textwidth, height =.28 \textheight]{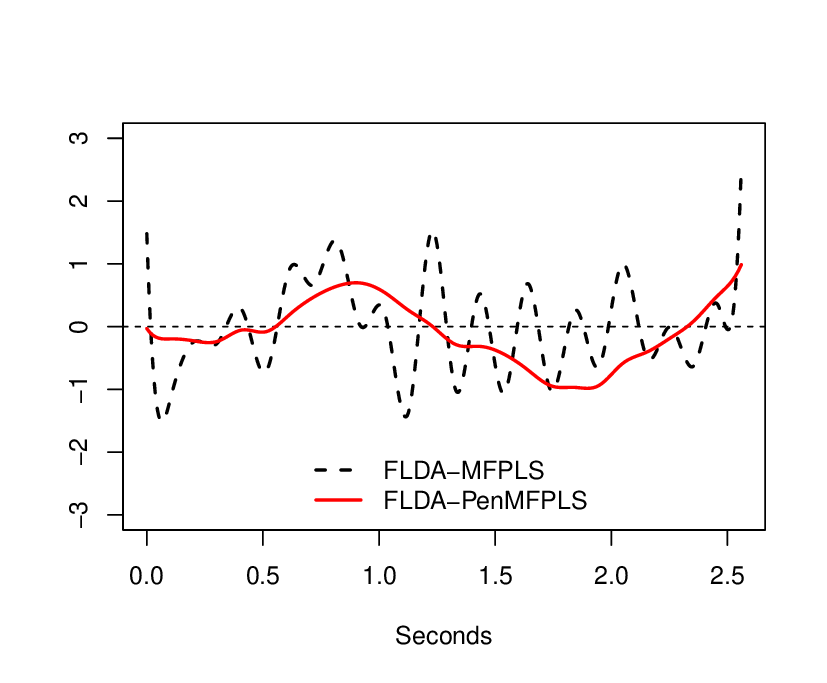}\\
\end{tabular}
\end{center}
\caption{Human activity data. Discriminant functions estimated by functional LDA based on non-penalized (dashed line) and penalized (solid line) multi-class functional PLS regression, using 4 and 7 PLS components, respectively.} \label{Fig:3}
\end{figure}
\end{center}

The discriminant functions estimated from penalized and non-penalized functional LDA can be seen in Figure \ref{Fig:3}. The necessity of smoothing in the estimation of the discriminant functions is obvious, with penalized functional LDA providing the smoothest and the most interpretable functions.

In order to evaluate the ability of penalized and non-penalized approaches for human activity pattern recognition, Table \ref{Tab:1} summarizes information related to the correct classification rate based on a leave-one-out cross-validation on the training sample ($CCR_{training}$), and based on the test sample classification ($CCR_{test}$). In this table, a multivariate version of LDA based on multi-class PLS regression (LDA-MPLS) is also considered for comparison purpose.
\begin{table}[ht]
\centering
\begin{tabular}{lccc}
& LDA-MPLS &  FLDA-MFPLS & FLDA-PenMFPLS  \\ \hline\noalign{\smallskip}
$CCR_{training}$ & 83\% & 95\% & 93\% \\
$CCR_{test}$ & 56\% & 67\% & \pmb{78\%} \\
$\# PLS cp's$ & 5 & 4 & 7\\
$\lambda$ & -& - & 0.41\\
\noalign{\smallskip}\hline\noalign{\smallskip}
\end{tabular}
\caption{Human activity data. Correct classification rate based on a leave-one-out cross-validation on the training sample ($CCR_{training}$) and based on the test sample classification ($CCR_{test}$), and the number of PLS components ($\# PLS cp's$) used in the functional LDA based on penalized and non-penalized multi-class functional PLS regression (FLDA-PenMFPLS and FLDA-MFPLS, respectively). $\lambda$ denotes the smoothing parameter of the penalized approach. A multivariate version of LDA based on multi-class PLS regression (LDA-MPLS) is also considered for comparison purpose.}
\label{Tab:1}
\end{table}

It can be seen that functional LDA performs better than the multivariate approach. Besides, between the two functional approaches (penalized and non-penalized), the one based on penalized multi-class functional PLS (FLDA-PenMFPLS) achieves the highest CCR in both, training and test samples. A considerable difference between the CCR in training and test samples highlights a possible overfitting mainly in the multivariate and in the non-penalized functional approaches. In the penalized approach this fact is not so remarkable, providing then the best classification rate.

\subsection{Gait data}

The gait data set comes from a wide experimental study developed in the Biomechanics laboratories of the Sport and Health Institute of the University of Granada (IMUDS). A total of 51 participants (25 boys and 28 girls) between 8 and 11 years old were involved in this study. In order to collect the data, twenty six reflective markers were placed on the children's skin. The kinematics data was recorded by a 3D motion capture system (Qualisys AB, G\"{o}teborg, Sweden). Each subject completed a cycle walking over the platform in three conditions (walking, carrying a backpack that weighs 20\% of the subject's weight and pulling a trolley that weighs 20\% of the subject's weight). For each subject, the 3-axial angular rotation were registered for each join (ankle, foot progress, hip, knee, pelvis, thorax) in all conditions. Finally, the angular rotation curve represents the observation for the subject's gait cycle for each join, axis direction and experimental condition. Interested readers could request access to the data from Jose M. Heredia-Jim\'enez and Eva Orantes-Gonz\'alez from IMUDS.

In this section, only a part of the above-described experimental data set has been considered. Exactly, we are interested on the thorax angular position (radians) measured on axis Z. This is a clear example of functional data with repeated measures in the sense that three curves, concerning the three experimental conditions, are available for each subject. In addition, each sample curve was recorded in 101 equidistant points of the gait cycle. So, the interval $[0, 100]$ represents the percentage of gait cycle completed by the subject.

\begin{center}
\begin{figure}[ht]
\begin{center}
\begin{tabular}{ccc}
\includegraphics[width=.30 \textwidth, height =.20 \textheight]{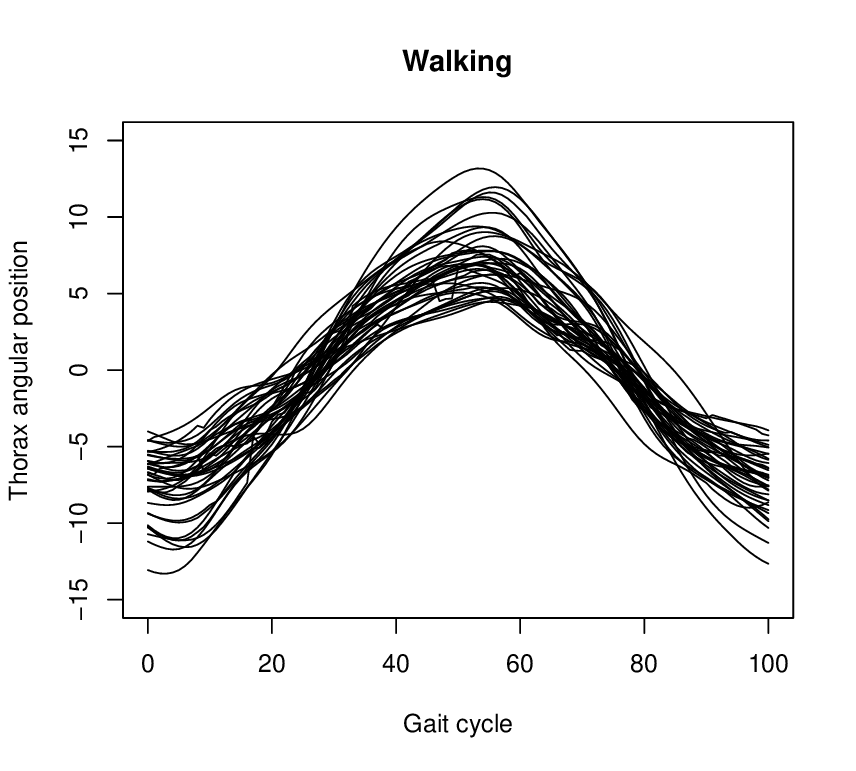}&
\includegraphics[width=.30 \textwidth, height =.20 \textheight]{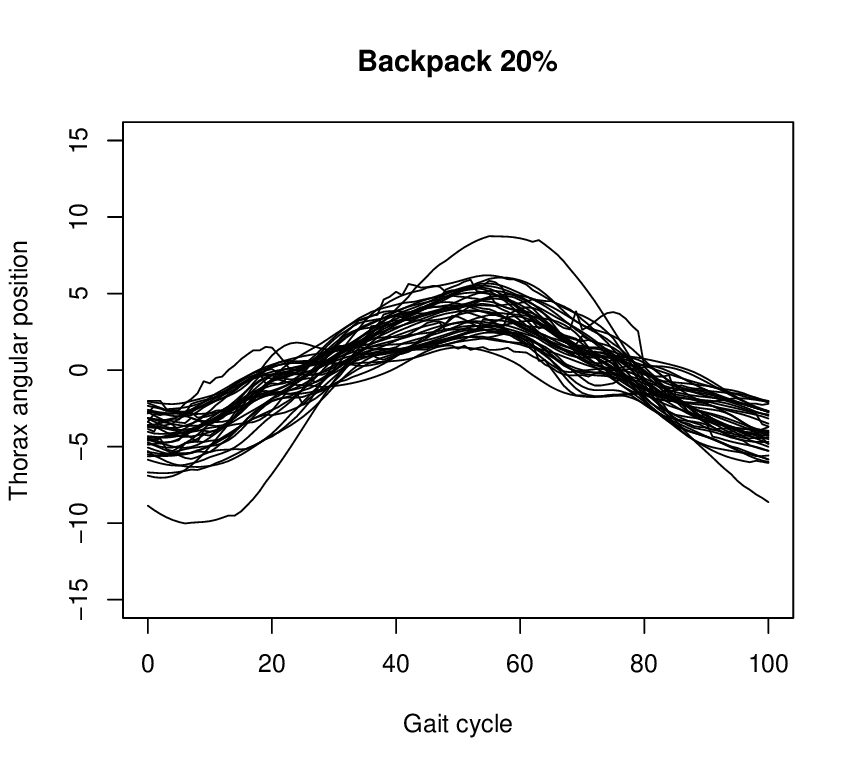}&
\includegraphics[width=.30 \textwidth, height =.20 \textheight]{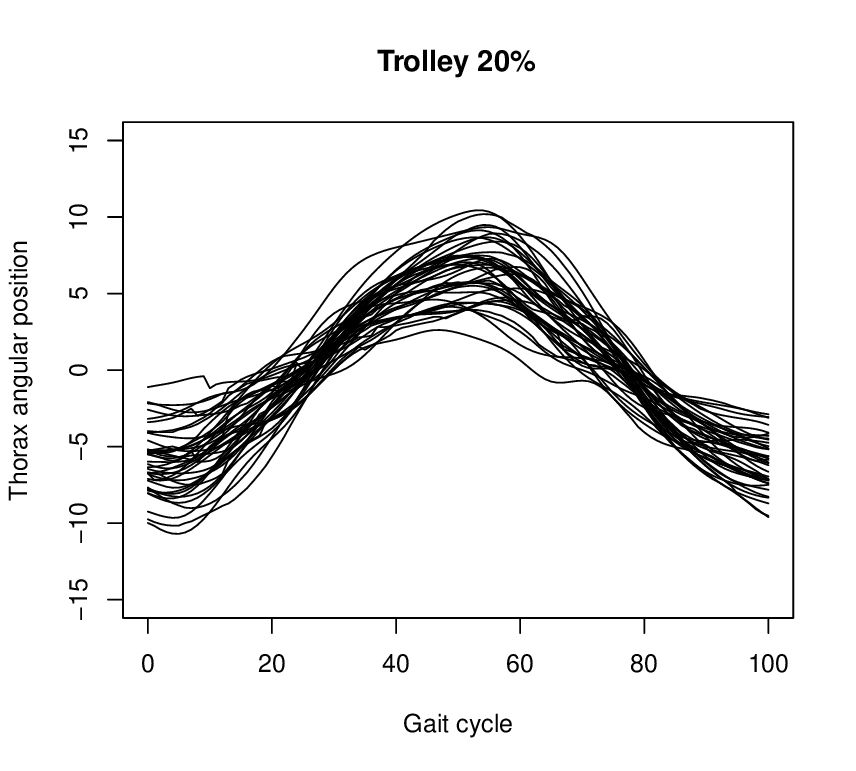}\\
\end{tabular}
\end{center}
\caption{Gait data. Raw data. Sample paths displayed separately by type of activity: walking, carrying a backpack and pulling a trolley, both weighing 20\% of the children's weight, from left to right, respectively.} \label{Fig:1_2}
\end{figure}
\end{center}

Subject 10 was removed from the sample for being an outlier. The remaining sample observations were distributed into training and test samples, with observations (repeated measures) related to 39 and 11 subjects, respectively. The raw data (displayed by experimental conditions) are shown in Figure \ref{Fig:1_2}.

\begin{center}
\begin{figure}
\begin{center}
\begin{tabular}{cc}
\includegraphics[width=.45 \textwidth, height =.28 \textheight]{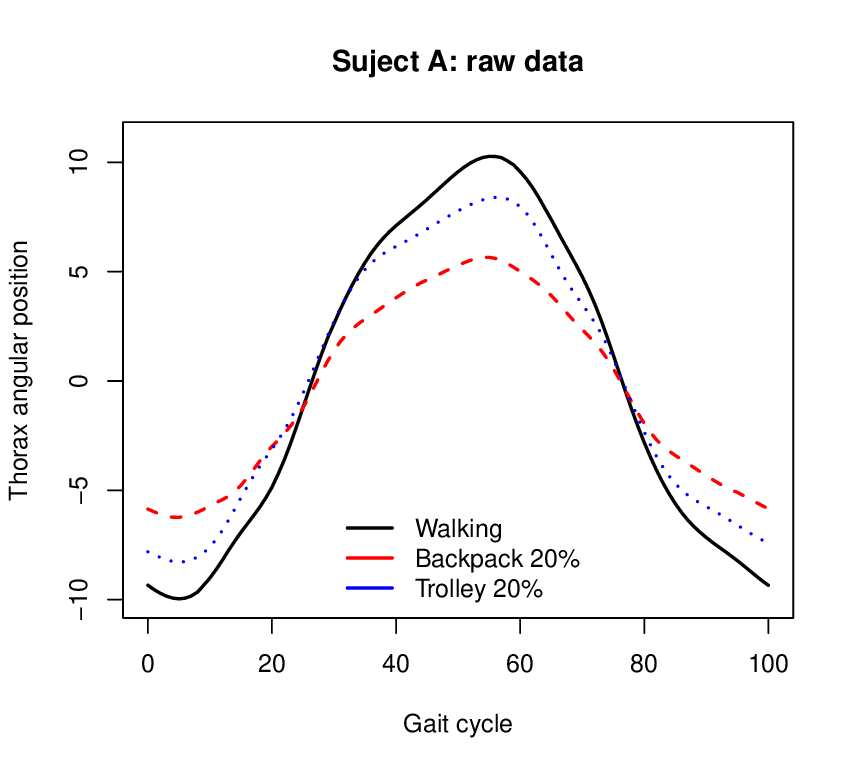}&
\includegraphics[width=.45 \textwidth, height =.28 \textheight]{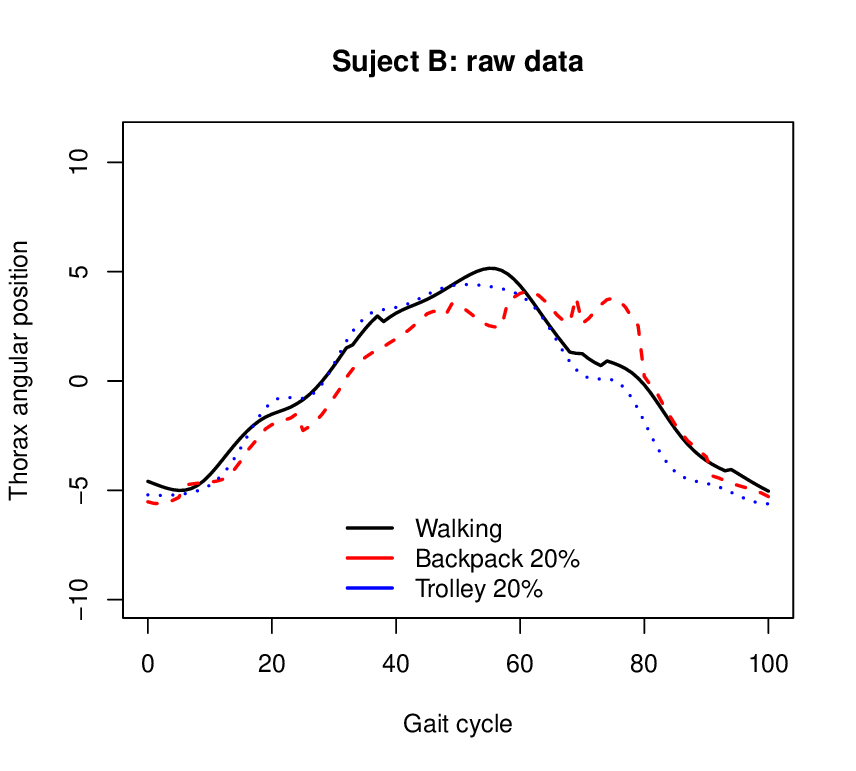}\\
\end{tabular}
\end{center}
\caption{Gait data. Raw data. Sample paths related to walking (solid line), carrying a backpack (dashed line) and pulling a trolley (dotted line), both weighing 20\% of the children's weight and walking downstairs from two sample individuals A and B, left and right panel, respectively.}
\label{Fig:1b_2}
\end{figure}
\end{center}

In order to visualize the within subjects and the between subjects variability, in Figure \ref{Fig:1b_2} the spectra related to the three gait events have been overlapped for two subjects A and B, left and right panel, respectively. It can be seen that walking and walking with a trolley provide sample paths with a more similar shape than walking with a backpack. Also it is interesting to highlight the between subjects variability (differences in scale and shape).

\begin{center}
\begin{figure}
\begin{center}
\begin{tabular}{ccc}
\includegraphics[width=.30 \textwidth, height =.20 \textheight]{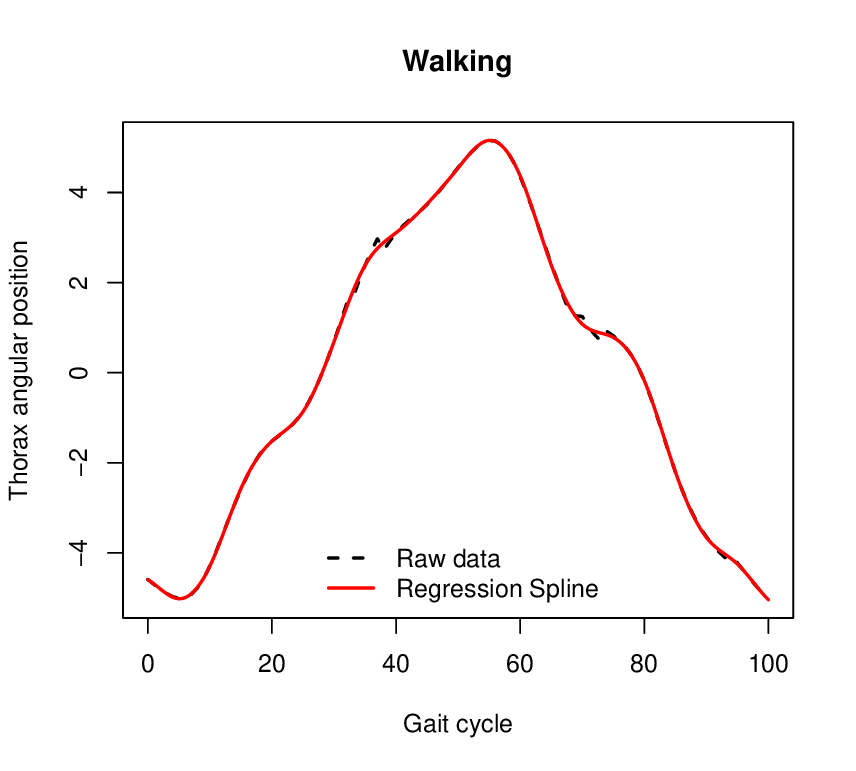}&
\includegraphics[width=.30 \textwidth, height =.20 \textheight]{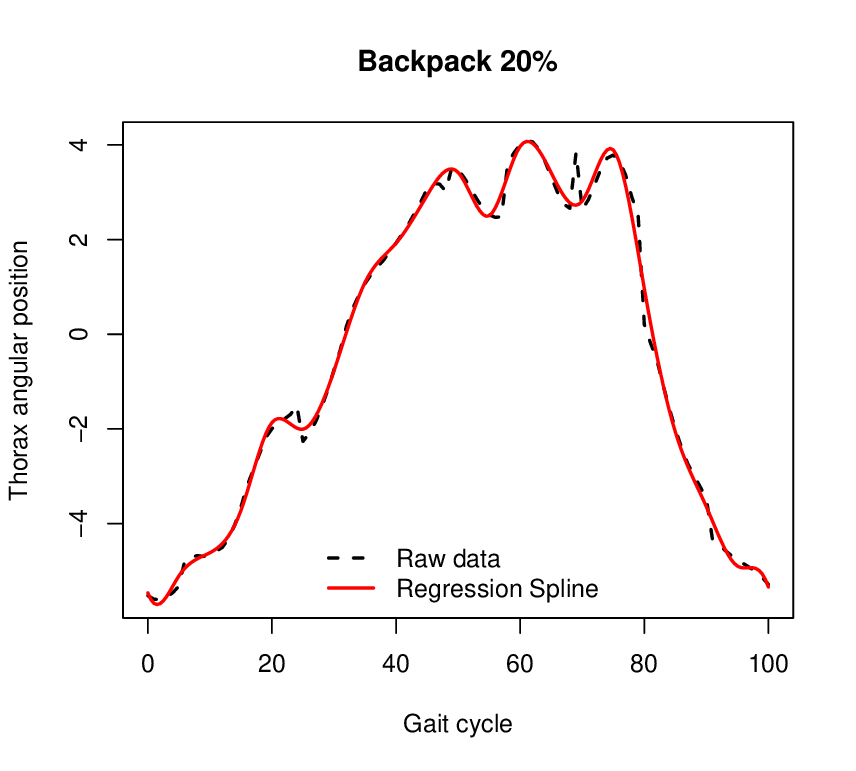}&
\includegraphics[width=.30 \textwidth, height =.20 \textheight]{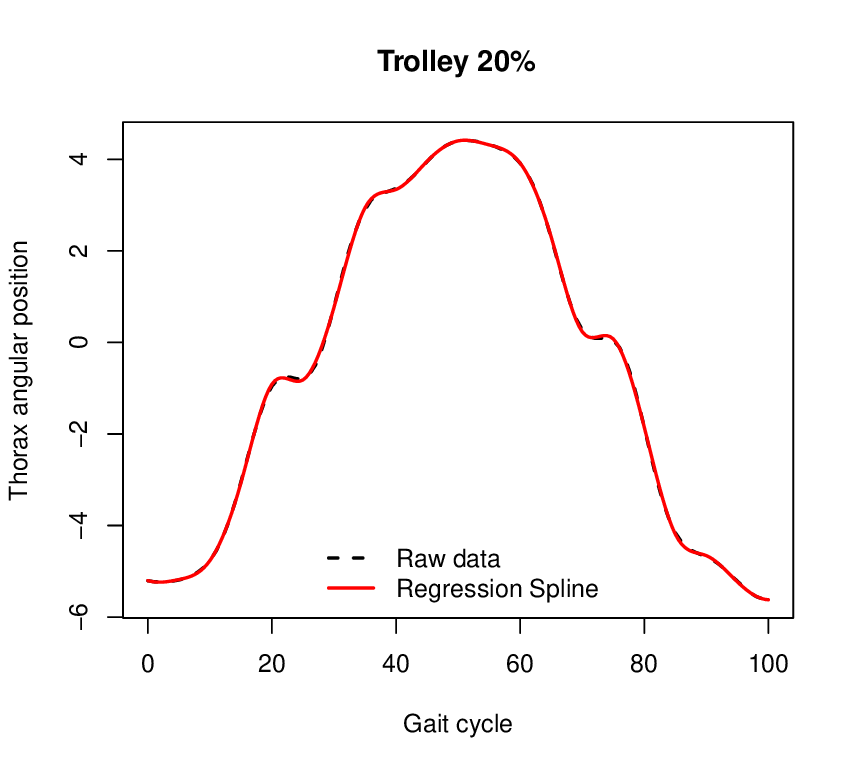}\\
\end{tabular}
\end{center}
\caption{Gait data. A sample path for each stimulus (dashed line) together with the regression spline (solid line) estimated with a cubic B-splines basis defined on 20 equidistant knots.} \label{Fig:2_2}
\end{figure}
\end{center}

\begin{center}
\begin{figure}
\begin{center}
\begin{tabular}{cc}
\includegraphics[width=.45 \textwidth, height =.28 \textheight]{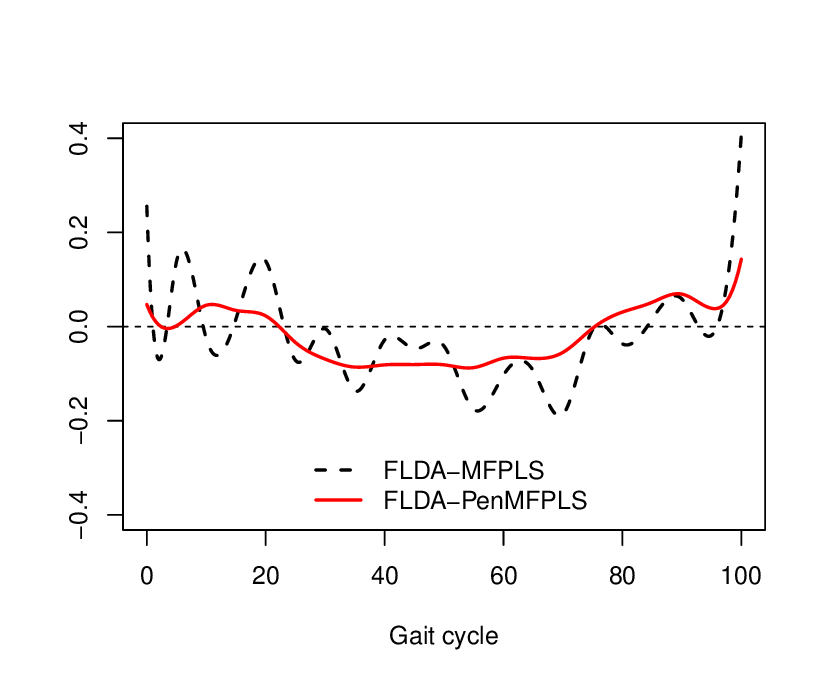}&
\includegraphics[width=.45 \textwidth, height =.28 \textheight]{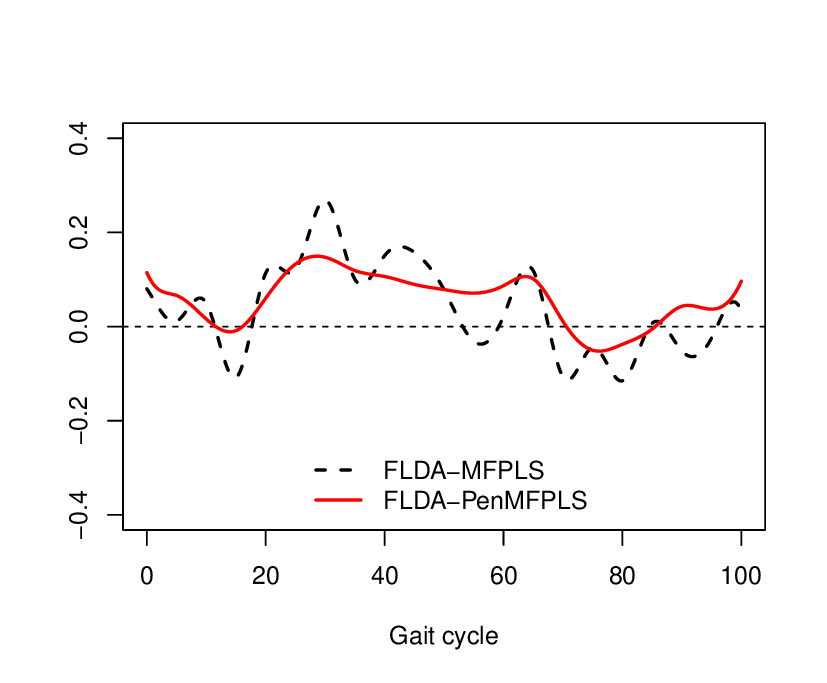}\\
\end{tabular}
\end{center}
\caption{Gait data. Discriminant functions estimated by functional LDA based on non-penalized (dashed line) and penalized (solid line) multi-class functional PLS regression, using 5 and 3 PLS components, respectively.} \label{Fig:3_2}
\end{figure}
\end{center}

The sample curves have been approximated by mean of a basis representation using a cubic B-spline basis defined on 20 equidistant knots. In Figure \ref{Fig:2_2} the raw data and the regression splines have been displayed. Once the sample curves have been approximated, the aim is the classification of the thorax angular rotation curves. In that sense, the discriminant functions estimated by penalized and non-penalized functional LDA can be seen in Figure \ref{Fig:3_2}. Once again the penalized approach provides the smoothest functions, providing an intuitive interpretation in relation with detecting periods in which the function is positive or negative.

The classification performance of the penalized and non-penalized approaches is shown in Table \ref{Tab:2_2}. It obvious that functional LDA performs better than the multivariate approach. Between the two functional approaches, the one based on penalized multi-class functional PLS (FLDA-PenMFPLS) achieves the best classification performance (the highest CCR on the test sample).
\begin{table}
\centering
\begin{tabular}{lccc}
& LDA-MPLS &  FLDA-MFPLS & FLDA-PenMFPLS  \\ \hline\noalign{\smallskip}
$CCR_{training}$ & 83\% & 87\% & 82\% \\
$CCR_{test}$ & 33\% & 61\% & \pmb{85\%} \\
$\#PLS cp's$ & 3 & 5 & 3\\
$\lambda$ &- & - & 2.87\\
\noalign{\smallskip}\hline\noalign{\smallskip}
\end{tabular}
\caption{Gait data. Correct classification rate based on leave-one-out cross-validation on the training sample ($CCR_{training}$) and based on the test sample classification ($CCR_{test}$), and the number of PLS components ($\#PLS cp's$) used in the functional LDA based on penalized and non-penalized multi-class functional PLS regression (FLDA-PenMFPLS and FLDA-MFPLS, respectively). $\lambda$ denotes the smoothing parameter of the penalized approach. A multivariate version of LDA based on multi-class PLS regression (LDA-MPLS) is also considered for comparison purpose.}
\label{Tab:2_2}
\end{table}

\section{Conclusions}
\label{conclus}

In this work a methodological solution to the problem of multi-class classification of functional data with repeated measures has been proposed. Functional linear discriminant analysis (FLDA) has been considered as a classifier, by solving the problem of infinite dimension of the functional data by means of a novel approach of functional PLS regression for repeated measures described in Section \ref{multi-class}. This work has been motivated by two real problems related to kinematic data. In both cases, data is affected by some noise, and then some type of penalization must be considered in the estimation of the discriminant functions. To solve this problem in the case of functional data with repeated measures, a multi-class approach for penalized functional PLS, that introduces a P-spline penalty in the definition of the inner product in the PLS algorithm, is proposed in Sections \ref{pls-pen} and \ref{ML-FPLS}.

In Sections \ref{sim} and \ref{results} the performance of the two functional approaches (penalized and non-penalized, FLDA-PenMFPLS and FLDA-MFPLS, respectively) are compared with a multivariate version of LDA based on multi-class PLS regression (LDA-MPLS). As we can see in Figure \ref{Fig:sim3} and Tables \ref{Tab:1} and \ref{Tab:2_2}, from a classification point of view, functional approaches perform better than the multivariate version, increasing the Correct Classification Rate (CCR) on a test sample in a very significant way. Regarding the two functional approaches, the one based on penalized multi-class functional PLS achieves the best classification ability, with a difference in the CCR of more than 10\% in both case studies.

The discriminant functions estimated from penalized and non-penalized functional LDA are shown in Figures \ref{Fig:3} and \ref{Fig:3_2}. It can be seen that the penalized version (FLDA-PenMFPLS) provides smoother functions, which are more interpretable in the sense of detecting periods in which the function is positive or negative.

Finally, it can be concluded that LDA based on a multi-class approach for penalized functional PLS is the most competitive method from both classification and estimation point of view.

\section*{Acknowledgements}
This research has been supported by the research projects $MTM2017-88708-P,$ Spanish Ministry of Economy and Competitiveness (also supported by the FEDER program) and $IJCI-2017-34038,$ Agencia Estatal de Investigacion, Ministerio de Ciencia, Innovacion y Universidades.

\bibliography{Biblio}

\end{document}